# **pyRMG**: A Framework for High-Throughput, Large-Cell DFT Calculations on Supercomputers

Ryan Morelock[1,*], Soumendu Bagchi[1], Emil Briggs[2], Wenchang Lu[2], Jerzy Bernholc[2,*], Panchapakesan Ganesh[1,*]

*[1]Center for Nanophase Materials Sciences, Oak Ridge National Laboratory, Oak Ridge, TN 37831, USA*

*[2]Department of Physics, North Carolina State University, Raleigh, NC, 27606*

*[*]Corresponding authors: Ryan Morelock (morelockrj@ornl.gov), Jerzy Bernholc (bernholc@ncsu.edu), Panchapakesan Ganesh ( ganeshp@ornl.gov)*

**Abstract**

Exascale computing delivers the raw power to simulate ever larger and more chemically realistic systems, but realizing this potential requires codes that can efficiently use thousands of processors. Our real-space multigrid (RMG) density functional theory (DFT) code's grid-decomposition approach scales nearly linearly with the number of GPUs, even for simulations exceeding thousands of atoms. This scalability makes RMG a compelling tool for high-throughput DFT studies of materials that would otherwise be bottlenecked in other codes (for example, by global Fast Fourier Transforms in plane-wave DFT). However, the limited workflow infrastructure for RMG has thus far constrained its adoption to a small user community. In this work, we present **pyRMG**, a Python package designed to streamline the setup and execution of RMG DFT calculations. Built on the pymatgen and ASE computational materials science Python packages, pyRMG automates input generation and convergence checking, and integrates with modern job schedulers (e.g., Flux) on leadership-class platforms such as *Frontier* and *Perlmutter*. We demonstrate pyRMG for a high-throughput study of strain effects in 2D 2L-$Bi_2Se_3$/2L-$NbSe_2$ heterostructures, which offers chemical insights into this system and shows that RMG-based workflows can converge with limited user intervention.

# I. Introduction

In recent years, the computational materials science community has developed a suite of open-access databases containing materials properties calculated from first principles using density functional theory (DFT), to aid in searching for new materials with desired properties. Prominent examples include the Materials Project[1], JARVIS[2], OQMD[3], NOMAD[4], AFLOW[5], and MatDB,[6] as well as numerous smaller, domain-specific platforms and datasets[7–10]. The standardization, transparency, and reproducibility of these repositories have enabled the characterization of structural[11,12], thermodynamic[13,14], and electronic properties[15,16] across hundreds of thousands of materials. These resources have significantly accelerated applications such as mapping synthesis pathways[17,18] and catalytic tuning[19], while laying the foundation for data-driven approaches, including machine-learned interatomic potentials[20–23].

Despite these successes, the computational expense of DFT calculations can lead to necessary compromises when generating these databases that limit dataset completeness and/or fidelity. Because datasets suitable for materials discovery and machine learning can require hundreds of thousands of calculations[24,25], most repositories rely on small unit cell representations to balance computational cost with dataset completeness, thereby excluding technologically-relevant systems that require large supercells—such as off-stoichiometric alloys or materials with low defect concentrations—and those needing larger cells to capture strong electronic correlations or long-range effects accurately[26,27]. Moreover, these databases predominantly contain experimentally stable, 3D bulk materials that are easier to converge, resulting in an underrepresentation of metastable, low-dimensionality materials, including surfaces and 2D homo- and heterostructures with complex stacking arrangements.

A contributing factor to this bias is that the computational frameworks underpinning most of the high-throughput, crystalline materials databases primarily interface with plane-wave DFT codes that are highly optimized for small periodic cells but perform less well for larger systems[28]. Even on highly parallel systems, performance bottlenecks can arise from the communication overhead of fast Fourier transforms (FFTs) required by plane-wave codes, serial or weakly parallel routines, or uneven load distribution within or among the nodes, where long runtimes cannot always be overcome by simply allocating more computational resources[28]. Such difficulties are compounded on modern CPU-GPU supercomputers with thousands of nodes: each node contains tens of CPU cores and multiple GPUs that dictate the floating-point performance. These challenges have effectively precluded high-throughput DFT studies of complex or extended material systems using models larger than ~1000 atoms, underutilizing the capabilities of modern high-performance computers (HPCs) such as *Perlmutter* at Lawrence

Berkeley National Laboratory, or exascale computers such as *Frontier* at Oak Ridge National Laboratory and *Aurora* at Argonne National Laboratory.

This paper describes the development of a Python-based interface, pyRMG, that wraps around the RMG (Real-space MultiGrid) electronic structure code[29,30] to support high-throughput calculations. RMG is an open-source, real-space electronic structure code whose grid-based representation is dual to a plane-wave basis, where finer grid spacing is commensurate to a higher plane-wave cutoff. RMG achieves equivalent accuracy compared to plane-wave codes[30], including VASP and Quantum Espresso, on the well-established Delta test for 71 elements in the periodic table[31]. Its real-space formulation avoids the extensive use of Fast Fourier Transforms (FFTs), which require whole domain communication and do not scale well across many nodes. RMG supports two eigenstate solvers: a Davidson subspace diagonalization, similar to what most plane-wave codes use, and a custom multigrid solver. The Davidson approach requires repeated global communication when constructing and diagonalizing the projected subspace Hamiltonian, which becomes increasingly costly as the number of electronic states grows. The multigrid algorithm uses a sequence of increasingly coarsened grids, starting from the base fine-level grid, to rapidly eliminate error components that oscillate at the coarsened-grid level, thereby accelerating convergence. The number of coarsened grid levels does not affect accuracy, but drastically reduces the number of iterations needed to reach convergence. Moreover, the multigrid solver treats each eigenstate locally, avoiding these global operations and achieving weakly-linear scaling across nodes for large systems, thereby accelerating convergence at extended length scales. RMG has been natively developed for clusters and supercomputers, and is therefore well optimized for parallel CPU-GPU architectures; real-space grids allow for efficient parallelization via domain decomposition and balanced workload distribution across nodes, utilizing all available GPUs and CPU cores. RMG runs efficiently on *Frontier* (AMD MI250X GPUs), *Aurora* (Intel Data Center Max 1550 GPUs) and *Perlmutter* (NVIDIA A100 GPUs) supercomputers, where it has been used for a variety of scientific applications, including as part of the QMCPACK Exascale Computing Project[32,33].

This work introduces pyRMG, a Python wrapper designed for users to more easily interface with the underlying RMG real-space DFT code on emergent Exascale computing resources. pyRMG's capabilities can promote efficient, high-throughput investigations that expand the scope of existing computational materials databases by leveraging RMG's accelerated convergence for systems with tens of thousands of electrons. pyRMG draws inspiration from Python toolkits such as pymatgen[34] and ASE (Atomic Simulation Environment)[35], which automate input generation and convergence monitoring for widely used DFT codes like VASP[36–38] and Quantum Espresso[39,40]. Similarly, pyRMG automates key components of RMG workflows, including wavefunction and k-point grid construction, SCF and force convergence checking, and the resubmission of calculations that fail or reach wall-time limits. Furthermore, it includes logic for the RMG-specific processor-grid assignment, automatically selecting node counts to reduce runtimes and prevent out-of-memory errors. pyRMG can also perform parallel, adaptive job scheduling using the MatEnsemble Python package[41], though MatEnsemble support is not strictly required to submit jobs with pyRMG. pyRMG is easily installed into Python or Conda environments on HPC clusters or local workstations supporting RMG, making it accessible to both existing and new RMG users; however, its lightweight design means that pyRMG is not as comprehensive as some other DFT workflow managers, like atomate2[42,43] for VASP, and it does not currently support features like multi-stage workflows or database hosting.

We demonstrate pyRMG's utility for high-throughput RMG DFT workflows by investigating the influence of strain and twist angle (film-on-substrate) on the relative stabilities and band gaps of 2D heterostructures of bilayer $Bi_2Se_3$ films on bilayer $NbSe_2$ substrates (2L-$Bi_2Se_3$/2L-$NbSe_2$). Van der Waals heterointerfaces between the topological insulator $Bi_2Se_3$[44] and the superconductor $NbSe_2$[45], which can host topologically protected surface states and potentially Majorana modes[46–49], make these materials intriguing candidates for topological qubits[50]. However, modeling these heterointerfaces can be challenging: they are 2D and require large supercells with sizable vacuum regions, and their DFT-predicted properties are likely sensitive to interfacial spacings, in-plane, substrate-film offsets, and the twist angle between $Bi_2Se_3$ and $NbSe_2$ layers at the interface. To investigate twist-angle effects, commensurate supercells between the film and substrate must be constructed. This can rapidly increase the total atom count and produce highly anisotropic cells, where some lattice vectors are much longer or shorter than others. The combination of large vacuum space, anisotropic cells, and varying atom (i.e., electron) numbers makes an investigation into twist-angle effects for 2L-$Bi_2Se_3$/2L-$NbSe_2$ a difficult validating workflow, challenging both the RMG code's ability to converge these demanding 2D cells and pyRMG's resource-allocation and processor-grid automation.

The remainder of the paper is organized as follows: Section II details the technical aspects of pyRMG and provides an overview of pyRMG's automated processor-grid assignment, which is sensitive to the available resources on the supercomputer target. We benchmark this approach with an example calculation, where user-supplied pyRMG flags modify the processor grid solution. Section III describes our dataset of lattice-matched, 2L-$Bi_2Se_3$/2L-$NbSe_2$ heterostructures, explains the supercell design metrics (lattice-match algorithm, von Mises strain, twist angle, and anisotropy), and provides the RMG parameters used for the workflow. Section

IV reports calculation runtimes, relates geometric strain to heterostructure energetics, and explores differences in stability and charge transfer between SCF and ion-optimized calculations, as well as those with spin-orbit coupling (SOC). Section V discusses other promising application domains for RMG + pyRMG, and Section VI summarizes and concludes our investigation.

## Technical Aspects

This section provides a technical overview of pyRMG executables that generate, manage and configure high-throughput RMG calculations for HPC platforms. This includes node-grid configuration to improve RMG calculation runtimes, as well as different modalities for serial (statically submitted) and parallel (dynamically spawned) workflows. Figure 1 illustrates how executables contained in pyRMG can be used to generate and submit RMG calculations in a high-throughput scheme.

### a. pyRMG Overview and MatEnsemble Interface

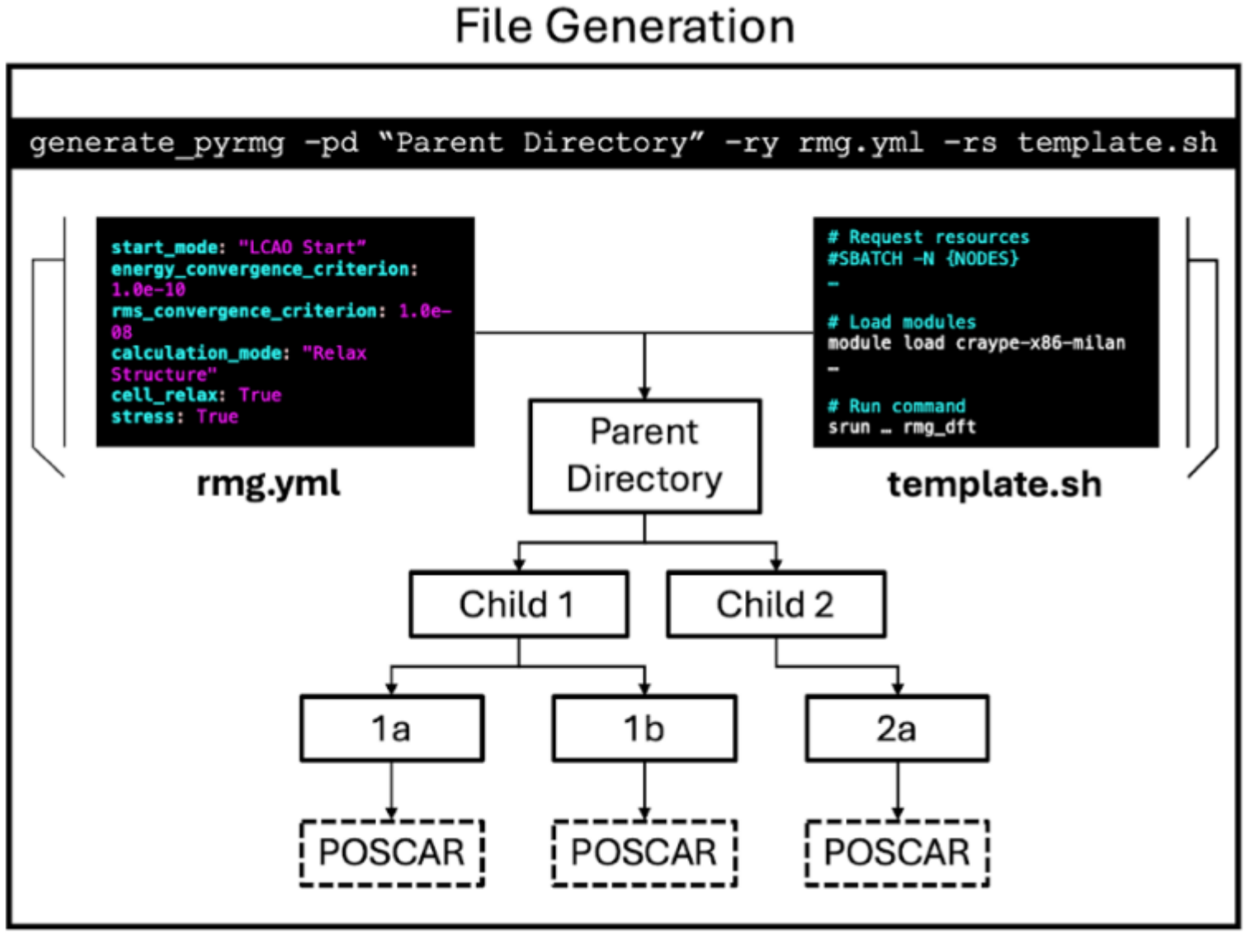

**Figure 1.** (a) Schematic of file generation using the `generate_pyrmg` command-line executable. The user specifies a parent directory containing POSCAR (or other compatible) structure files, an rmg.yml configuration file, and a submission template; these inputs are used to generate the corresponding RMG input files and single-job shell submission scripts. (b) Example of an RMG job submission directory populated with rmg_input files and submission scripts by `generate_pyrmg`. The `submit_pyrmg` executable scans all child directories for un-converged forcefield.xml files (generated by the RMG binary) and resubmits individual jobs in these directories.

pyRMG, which depends on the ASE and pymatgen Python packages, contains logic to generate, configure, and submit RMG calculations. The `config_pyrmg` executable allows the user to set defaults values for their HPC platform, such as the path to the RMG executable (binary), the pseudopotential directory path, and resource flags such as the CPUs or GPUs per node. The `generate_pyrmg` executable scans a user-specified directory tree for files readable by pymatgen's Structure object (including VASP POSCAR and .cif files) and generates RMG input files based on configured defaults and user-supplied arguments, such as the YAML directive file. After input file generation, jobs can be submitted with either the `submit_pyrmg` or the MatEnsemble-aware `matsemble_pyrmg` executable. Both job submission executables inspect convergence flags in the RMG input and the forcefield.xml output to decide whether a run has converged (and can be skipped) or should be (re)submitted. Additional implementation details, including job submission templates (see Figure 10 and Figure 11), are provided in Appendix A.

Job-chaining with an HPC job scheduler (e.g., Slurm) is one way to wrap multiple RMG runs into a single submission, and we provide an example script that uses Slurm job-chaining on the pyRMG GitHub. However, chained jobs are inflexible and ill-suited for heterogeneous DFT workflows, where jobs require different resources and have different convergence times. To give users greater control over their workflows, `matsemble_pyrmg` integrates pyRMG with the highly asynchronous MatEnsemble task manager, which provides a Python interface to the Flux job scheduler[51]. Each individual RMG run becomes a Flux task ("fluxlet") managed by a SuperFlux manager object, which dynamically spawns and monitors fluxlets using the available pool of resources, i.e., CPU-GPU nodes for GPU-supported RMG. The manager object replaces finished tasks with tasks that have not yet started, and the resource pool remains saturated until the full workflow completes. This type of dynamic, multi-resource scheduling is particularly advantageous on systems prioritizing large-node jobs with short wall times, like ORNL's *Frontier*.

### b. **Processor Grid Generation**

RMG is most efficient when the real-space meshes covering simulation cells are partitioned into uniform subdomains across all processing elements, as poor grid choices can cause load imbalance and calculation slowdowns or even failures. The base RMG code can auto-generate processor grids for the CPU-GPU resources available to it at runtime, but its heuristical approach often produces inhomogeneous partitions for highly anisotropic cells (e.g., 2D heterostructures with elongated lattice vectors). Expecting the user to manually construct grids for high-throughput workflows, or for highly anisotropic cells, is impractical, so pyRMG automatically solves for processor grids based on two user-supplied tags: an "electrons-per-GPU" target and a grid-divisibility exponent, $n$.

The "electrons-per-GPU" target defines an ideal, total GPU count ($G_{ideal}$) for a simulation cell based its valence electrons, which are pseudopotential-specific. pyRMG can currently calculate $G_{ideal}$ for SG15 ONCV pseudopotentials included with the base RMG distribution and user-supplied pseudopotentials with the .upf format:

$$G_{ideal} = \frac{Total\ valence\ electrons}{Electrons\ per\ GPU} \quad (1)$$

Fixing a GPU target based on the number of valence electrons in the cell, which helps define the total number of wavefunctions used by the calculation, can avoid memory overflow errors when too few GPUs are allocated for a single job. It also promotes more consistent timings per self-consistent field (SCF) step across cells with widely varying atom counts, which can be useful when packaging multiple jobs into a single, high-throughput workflow with a fixed wall time.

The grid-divisibility exponent, $n$, controls a simulation cell's processor-subdomain sizes so that each subdomain meets RMG's finite-difference minimum of at least kohn_sham_fd_order/2 grid points per direction; RMG's default kohn_sham_fd_order is 8, but this is adjustable between 6 and 12. In practice, the minimum number of grid points is more often achieved when the processor-subdomain dimensions are divisible by larger powers of 2 ($2^n$, oftentimes with $n \geq 3$). Increasing $n$ helps avoid grid-point errors, which can become particularly problematic for anisotropic meshes, while decreasing it gives the user greater flexibility in the total processor (GPU) count, which can be useful depending on the HPC system (e.g., system architecture, available allocation).

pyRMG solves for a simulation cell's optimal processor grid based on the "electrons-per-GPU" target and grid-divisibility exponent values passed to the `generate_pyrmg` executable as --electrons_per_gpu and --grid_divisibility_exponent flags. Details of the scoring function used to rank candidates and select the optimal grid are provided in Appendix A.

### c. Timing Benchmarking

To quantify the effect of processor grid choice on RMG calculation timing, we compare average SCF step times (seconds, s) for different grids auto-generated by pyRMG for Nb (110) slab calculations. The slab contains six Nb atoms (78 valence electrons) and 20 Å of vacuum space (lattice parameters a = 3.259 Å, b = 4.609 Å, and c = 27.654 Å), matching the vacuum spacing used in our 2D heterostructure investigation. RMG "kpoint_mesh" and "kpoint_distribution" flags were fixed to [10, 7, 1] and 4, respectively, for all single point calculations. The "cutoff" parameter of 300 Rydberg passed to pyRMG via our YAML directive file generates an RMG wavefunction grid of [34, 48, 288]; this dense wavefunction grid allows more variable processor grids to be solved for by pyRMG, as listed in **Table 1**.

**Table 1** shows that pyRMG's auto-generated grid of [1, 1, 8] reduces the average SCF step time by 13% compared to RMG's internally constructed grid of [2, 2, 2], despite using the same total number of GPUs (8). Although SCF steps become faster as the GPU count increases, these gains diminish rapidly, yielding an optimal balance between resource usage and runtime at --electrons_per_gpu=4. Notably, the --electrons_per_gpu=1 case requesting 408 GPUs fails, a limiting condition where individual subdomains become too small to support valid decomposition. Our benchmarking demonstrates that using pyRMG to generate processor grids, instead of the default RMG grid solver, reduces SCF runtimes, which we expect will provide meaningful savings for high-throughput workflows. A step-by-step example solving the optimal processor grid for Nb (110) is included in Appendix A.

**Table 1.** Comparison of SCF step times for Nb (110) slab single-point ("Quench Electrons") calculations as a function of --electrons_per_gpu. Processor grids solved by pyRMG are reported in the $\{p_x, p_y, p_z\}$ column.

| `--electrons_per_gpu` | $\{p_x, p_y, p_z\}$ | Total GPUs | Mean time per SCF step (s) |
|---|---|---|---|
| 1 | [2, 3, 17] | $4 \times 102 = 408$ | Error |
| 2 | [2, 2, 12] | $4 \times 48 = 192$ | 2.70 |
| 3 | [1, 2, 11] | $4 \times 22 = 88$ | 3.00 |
| 4 | [1, 2, 10] | $4 \times 20 = 80$ | 2.94 |
| 5, 6, 7 | [1, 2, 9] | $4 \times 18 = 72$ | 3.82 |
| 8 and higher | [1, 1, 8] | $4 \times 8 = 32$ | 5.67 |
| ***RMG Default*** | [2, 2, 2] | $4 \times 8 = 32$ | 6.48 |

## II. Heterostructure Construction and Analysis

This section describes how 2L-$Bi_2Se_3$/2L-$NbSe_2$ heterostructure calculations were generated, downselected and initialized for our high-throughput RMG + pyRMG investigation.

We provide an overview of lattice-match, the heteroepitaxy algorithm developed by Zur & McGill[52]. Our own HeteroBuilder Python package (part of the Q-CAD platform) uses lattice-match (as implemented by pymatgen's CoherentInterfaceBuilder) to construct 2L-$Bi_2Se_3$/2L-$NbSe_2$ heterostructures with user-supplied vdW-layers and interfacial spacings. We also define terms like the von Mises strain, twist angle, and grid anisotropy, which are needed for our DFT analysis in Section IV, and detail the RMG DFT settings used at different points in the workflow. Figure 2 provides a graphical overview of the workflow employed in this investigation.

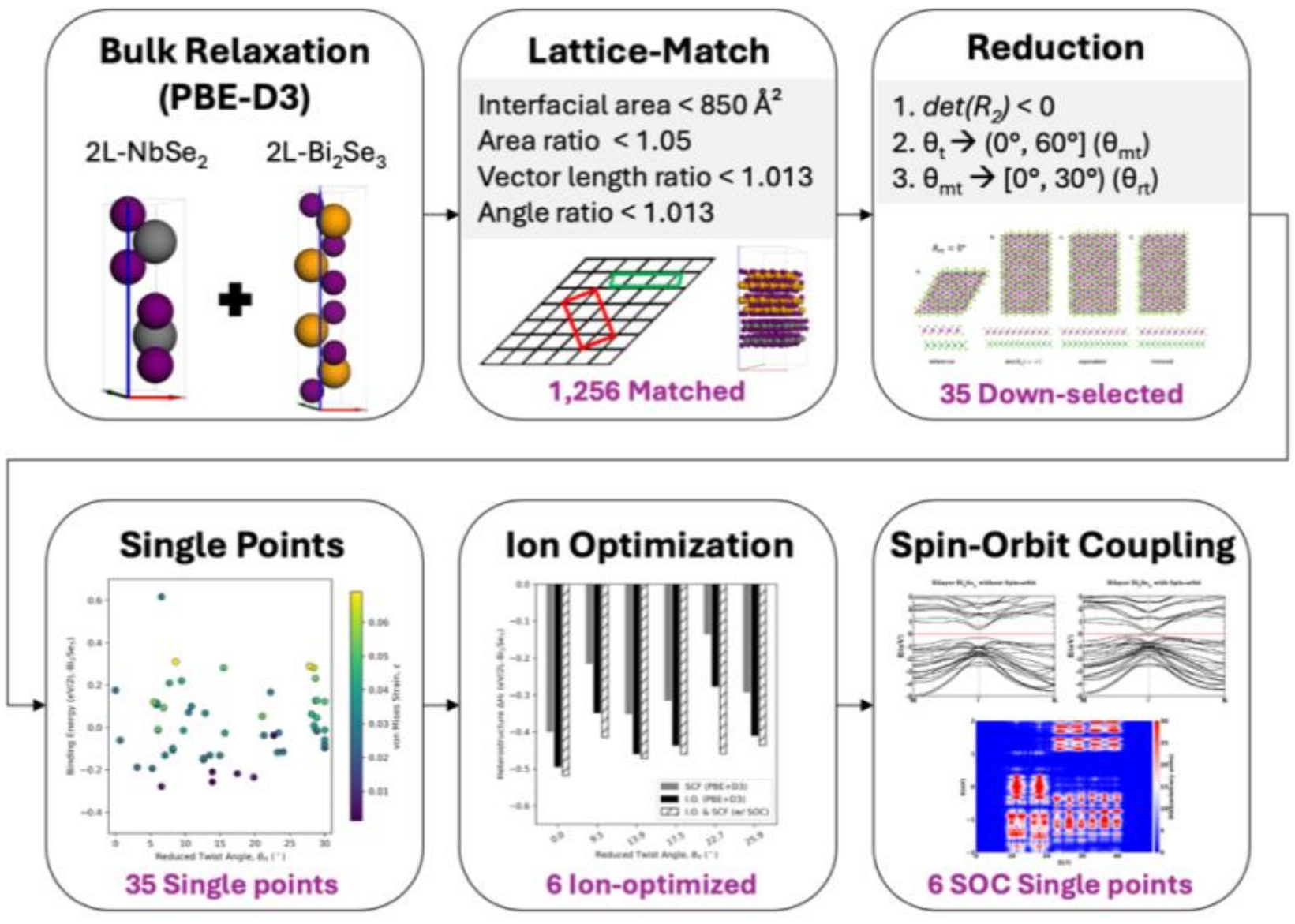

**Figure 2.** Workflow for the 2L-$Bi_2Se_3$/2L-$NbSe_2$ heterointerface study. Bulk $NbSe_2$ and $Bi_2Se_3$ structures were first optimized, and two-layer vdW slabs were lattice-matched to generate 1,256 candidate heterostructures. Configurations with inverted interfaces were removed for consistency, and the remaining structures were grouped by their reduced twist angle, $\theta_{rt}$, yielding 35 unique systems containing 168–1,046 atoms. All 35 systems were SCF converged, and 6 representatives were ionically-optimized and SCF converged with spin-orbit coupling (SOC).

## a. Background and Metrics

### *i. Overview of the lattice-match algorithm*

Lattice-match solves for commensurate superlattices between two different 2D lattices, e.g., film and substrate slabs, within user-specified geometric tolerances, including matched area, lattice vectors lengths, and internal angle. Its approach limits coordination environment distortions in the matched film and substrate cells, providing a good first approximation for heteroepitaxial growth that can offer reasonable (near-equilibrium) geometries for DFT calculations. This algorithm has four steps:

1. Solve for the primitive lattice vectors.

Reduce the 2D lattice vectors defining the film $(a, b) \in \mathbb{R}^2$ and substrate $(a', b') \in \mathbb{R}^2$ slabs to primitive lattice vectors $a_0, b_0$ and $a_0', b_0'$ using the iterative approach outlined by Zur & McGill. While this approach can solve for different primitive vectors depending on the starting lattices, the primitive vector magnitudes ($\|a\|, \|b\|$) and the angle between them ($\alpha$) are uniquely defined.

2. Find nearly equivalent superlattice areas.

Compute the primitive cell areas $|a_0 \times b_0|$ and $|a_0' \times b_0'|$ and find integer multiples $(N_{film}, N_{sub})$ that match within a user tolerance, providing candidate areas for the film and substrate.

$$A_{film} = N_{film} \times |a_0 \times b_0| \approx A_{sub} = N_{sub} \times |a_0' \times b_0'| \tag{2}$$

A user-supplied area-ratio tolerance sets the upper limits of $N_{film}$ and $N_{sub}$.

3. Construct candidate superlattices.

Solve for superlattice vectors $(u, v)$ and $(u', v')$ with candidate areas $A_{film}$ and $A_{sub}$ determined by $N_{film}$ and $N_{sub}$:

$$\begin{pmatrix} u \\ v \end{pmatrix} = \begin{pmatrix} i & j \\ 0 & m \end{pmatrix} \begin{pmatrix} a \\ b \end{pmatrix} \tag{3}$$

Where the integers $i, j,$ and $m$ defining the superlattice transformation matrix are subject to the conditions $i \cdot m = N$, $i, m > 0$, and $0 \leq j \leq m - 1$, and provide the complete set of possible superlattices.

4. Test commensurability and accept within tolerances.

Reduce each $(u, v)$ and $(u', v')$ to the primitive lattice vectors $u_0, v_0$ and $u_0', v_0'$ using the approached outlined in Step 1. If $\|u_0\| \approx \|u_0'\|$, $\|v_0\| \approx \|v_0'\|$ and $\alpha_0 \approx \alpha_0'$ within user-supplied lattice vector length and angular tolerances, the film and substrate supercells defined by $(u, v)$ and $(u', v')$ are accepted as commensurate lattices.

*ii. von Mises strain, $\varepsilon_{vM}$*

Heterolattices created from larger film and substrate superlattices (Step 2.) generally have smaller area offsets than heterostructures created from smaller superlattices. Zur & McGill show, for example, that as the area-matched, scalar-integer multiples of CdTe/GaAs primitive lattices increased, the area differences in the constructed heterointerfaces decreased[52]. Smaller area mismatches require less local deformation of the film relative to the substrate and hence induce less *geometric* strain.

The deformation tensor for lattice-matched superlattice vectors is computed as follows: for the 2D film and substrate superlattices constructed in the previous section, $(u, v) \in \mathbb{R}^2$ and $(u', v') \in \mathbb{R}^2$, the normal is:

$$n = u \times v, \quad n' = u' \times v' \tag{4}$$

Scaling the substrate vector's magnitude to the film vector's magnitude avoids introducing artificial out-of-plane stretch when solving the transformation matrix between the two lattices

$$\tilde{n}' = n' \frac{\|n\|}{\|n'\|} = n \tag{5}$$

which allows the 3x3 bases to be built as column matrices:

$$A_{film} = [u, v, n], \quad A_{sub} = [u', v', n] \tag{6 − 7}$$

The linear map $F$ between the film and substrate bases can be expressed as:

$$A_{film} F = A_{sub}, \quad F = A_{sub} A_{film}^{-1} \tag{8 − 9}$$

In general, $F$ can be decomposed into the 3D rotational, $R$, and symmetric, positive-definite right stretch, $U$, matrices. Under polar decomposition, $F = RU$, and thus $C = F^T F = (RU)^T (RU) = U^T R^T R U = U^2$, as $R^T R = 1$ and $U^T U = U^2$ by the matrix identity ($U$ is symmetric). Therefore, the Green-Lagrange strain, $E$, only depends on stretch $U$.

$$E = \frac{1}{2}(F^T F - I_3) = \frac{1}{2}(U^2 - I_3) \tag{10}$$

The von Mises (equivalent) strain scalar can then be computed from the deviatoric strain $E' = E - \frac{1}{3} tr(E) I_3$, which is a square matrix

$$\varepsilon_{vM} = \sqrt{\frac{2}{3} tr((E')^2)} = \sqrt{\frac{2}{3} \sum_{i.j} \varepsilon'_{ij} \cdot \varepsilon'_{ij}} \tag{11}$$

where only in-plane deformations (nonzero $E_{xx}, E_{yy}, E_{xy}, E_{yx}$) are present by construction.

Larger $\varepsilon_{vM}$ correspond to greater film deformation and typically larger energetic penalties. By relaxing the maximum interfacial-area constraint during lattice matching, larger supercells that reduce this deformation can be obtained, thereby reducing geometric distortion and yielding more stable interfaces. Accurately evaluating the *ab initio* energetics and other properties of these minimally-strained, yet often large and irregularly-shaped, heterostructures motivates the use of scalable DFT codes such as RMG with automated setup tools like pyRMG.

### *iii. Twist angle, $\theta_t$*

The linear map $F$ contains both rotation $R$ and stretch $U$ components, and we use a polar decomposition approach to solve for $R$ and define the twist angle of the film with respect to the substrate. The twist angle between film and substrate can strongly affect 2D electronic properties (e.g., magic-angle flat bands in twisted bilayer graphene[53] or localized excitons in transition metal dichalcogenide moiré heterobilayers[54]) and is expected to be an important design consideration for our 2L-$Bi_2Se_3$/2L-$NbSe_2$ heterostructures.

Previously, we built full 3x3 bases $A_{film} = [u, v, n]$ and $A_{sub} = [u', v', n]$. Because the out-of-plane normals of $A_{film}$ and $A_{sub}$ are equal, the principal rigid-body rotation lies in the xy-plane, and the twist angle can be extracted from the in-plane, 2x2 rotation matrix.

As shown when computing the von Mises strain, the linear map $F \in \mathbb{R}^3$ between the film and substrate bases can be decomposed into rotation and stretch components:

$$F = A_{sub}A_{film}^{-1} = RU \tag{12}$$

Where $F^TF = U^2$, or $U = (F^TF)^{1/2}$. The rotation matrix $R$ is therefore equal to:

$$R = A_{sub}A_{film}^{-1}U^{-1} = A_{sub}A_{film}^{-1}\left((F^TF)^{\frac{1}{2}}\right)^{-1} \tag{13}$$

The in-plane block, $R_{2x2}$, is a function of the rotation angle $\theta$ in the xy-plane, which defines our twist angle:

$$R_2 = \begin{pmatrix} R_{xx} & R_{xy} \\ R_{yx} & R_{yy} \end{pmatrix} = \begin{pmatrix} cos\theta & -sin\theta \\ sin\theta & cos\theta \end{pmatrix} \quad (14)$$

$\theta$ can be solved as $\theta = \arctan(R_{yx}, R_{xx}) = \arctan(sin\theta, cos\theta)$. However, arctan unambiguously maps $\theta$ to the interval $\left(\frac{-\pi}{2}, \frac{\pi}{2}\right)$, as values in the first and fourth quadrants ($\frac{y}{x} = \frac{-y}{-x}$) and second and third quadrants ($\frac{y}{-x} = \frac{-y}{x}$) are equivalent. Instead, we can generally solve the twist angle as $\theta_t = \arctan 2(R_{yx}, R_{xx})$, which imposes a half-turn ($\pm x$) on the $\arctan(R_{yx}, R_{xx})$ solution when x < 0, placing all points in their correct quadrants and mapping $\theta$ to the range $(-\pi, \pi]$.

Because pymatgen's CoherentInterfaceBuilder lattice-matching routine returns $A_{film}$ and $A_{sub}$ bases with differing signs or orderings, polar decomposition can yield improper rotation matrices with $\det(R_2) = -1$ instead of $\det(R_2) = 1$ (see b. Dataset Generation). This represents a reflection × rotation in the bases instead of a pure rotation. Twist angles are not directly comparable between $\det(R_2) = -1$ and $\det(R_2) = 1$, and to ensure consistency in this investigation we limited ourselves to structures with $\det(R_2) = 1$.

### *iv. Processor Grid Anisotropy, $\mathcal{A}$*

Zur's algorithm places no explicit geometric restrictions on the matched heterointerfaces, so the resulting supercells can have very different in-plane lattice lengths $a$ and $b$ and interior angles $\gamma$ far from 90°. This can generate many different matched lattices with different twist angles, including highly elongated or strongly anisotropic cells that can pose problems for charge mixing techniques, slowing DFT convergence. Adjusting the processor grid used by RMG (see Sections

II b/c) can mitigate this somewhat, but SCF convergence is expected to slow as the cell anisotropy (and by extension, the solved processor grid) becomes more variable.

To quantify this, we first define a simple processor-grid anisotropy metric, $\mathcal{A}$. Let $w_x, w_y, w_z$ be the processors along the Cartesian coordinate directions x, y and z, and let $\bar{w} = (w_x + w_y + w_z)/3$. Then the processor grid anisotropy metric, $\mathcal{A}$, that measures the root-mean-square deviation of the three grid directions to the mean grid is

$$\mathcal{A} = \sqrt{\frac{(w_x - \bar{w})^2 + \left(w_y - \bar{w}\right)^2 + (w_z - \bar{w})^2}{3}} \tag{15}$$

For 2D heterostructures a large fraction of the c-axis often contains vacuum, so $w_z$ overestimates the "active" grid. We therefore use an effective z-extent that ignores grid in the vacuum space:

$$w_z^{eff} = w_z\left(1 - \frac{\Delta_{vac}}{\|c\|}\right) \tag{16}$$

Where $\Delta_{vac}$ is the vacuum thickness along the c-vector (which has only z components in this study) and $\|c\|$ is its magnitude. Replacing $w_z$ with $w_z^{eff}$ for $\mathcal{A}$ yields an anisotropy metric that only considers the region with atomic sites, where the wavefunction is non-negligible.

Larger $\mathcal{A}$ indicates a more uneven (anisotropic) processor grid, and which we found strongly correlates with longer SCF times for our preliminary set of 91 structures (Appendix B).

## b. Dataset Generation

We first relaxed the lattices and site positions of tetragonal $Bi_2Se_3$ (15-atom primitive cell, 3 vdW layers) and hexagonal $NbSe_2$ (6-atom primitive cell, 2 vdW layers) with RMG DFT. The technical components of RMG DFT ionic relaxations with pyRMG, including handling grid-incompatibility errors before resubmission, are provided in Appendix A. Two vdW layers of the bulk-relaxed $Bi_2Se_3$ were used as the heterostructure film (2L-$Bi_2Se_3$), and two vdW layers in the bulk-relaxed $NbSe_2$ were used as the substrate (2L-$NbSe_2$). Heterointerfaces were constructed using HeteroBuilder, a standalone package we developed manipulate 2D materials, which uses pymatgen's CoherentInterfaceBuilder to find commensurate supercells via lattice-match.

To restrict heterostructure cell sizes and von Mises strains, we limited the interfacial area to < 850 Å$^2$, the area ratio to < 1.05, and the vector length and angle ratios to < 1.013. These settings constrained the largest solved cells to < 1,050 atoms and the maximum von Mises strain to < 0.016. We only considered interlayer vdW slab terminations (no intralayer terminations for 2L-$Bi_2Se_3$ and 2L-$NbSe_2$), set the vacuum space to a 20 Å for all heterostructures, and fixed the spacing between the films and the substrates to 3.2 Å. We chose 3.2 Å as the interfacial distance based on RMG DFT energy-vs-distance scans computed for two representative heterostructures, which are reported in Figure 12. These cells were only used to estimate the exfoliation energy, and were not otherwise included in our strain-restricted dataset or analyzed.

With these constraints, HeteroBuilder initially returned 1,256 candidate heterostructures. We first excluded structure mappings with improper rotation matrix determinants, $\det(R_2) < 0$, before grouping the remaining candidates by reduced twist angle, $\theta_{rt}$. $\theta_{rt}$ accounts for the six-fold symmetry of $Bi_2Se_3$ and $NbSe_2$ monolayers by taking the modulo $60°$, $\theta_{mt}$, which maps each raw twist angle $\theta_t$ to the interval $(0, 60°]$. Our preliminary results in Appendix B show a mirror symmetry in formation energies about $\theta_{mt} = 30°$, where structures with the same $\varepsilon_{vM}$ have the same formation energy, $\Delta H_f$ (e.g., structures with $\theta_{mt} = 40°$ and $\theta_{mt} = 20°$ are the same). This does not necessarily mean that the structures in each reduced group are exactly equivalent (see a. vs d. in Figure 3), but is a reasonable justification to group structures by their mirrored angles, $\theta_{rt}$. Our dataset contains the smallest representative heterostructure from each group, comprising 35 structures with supercells ranging from 168 to 1,046 atoms.

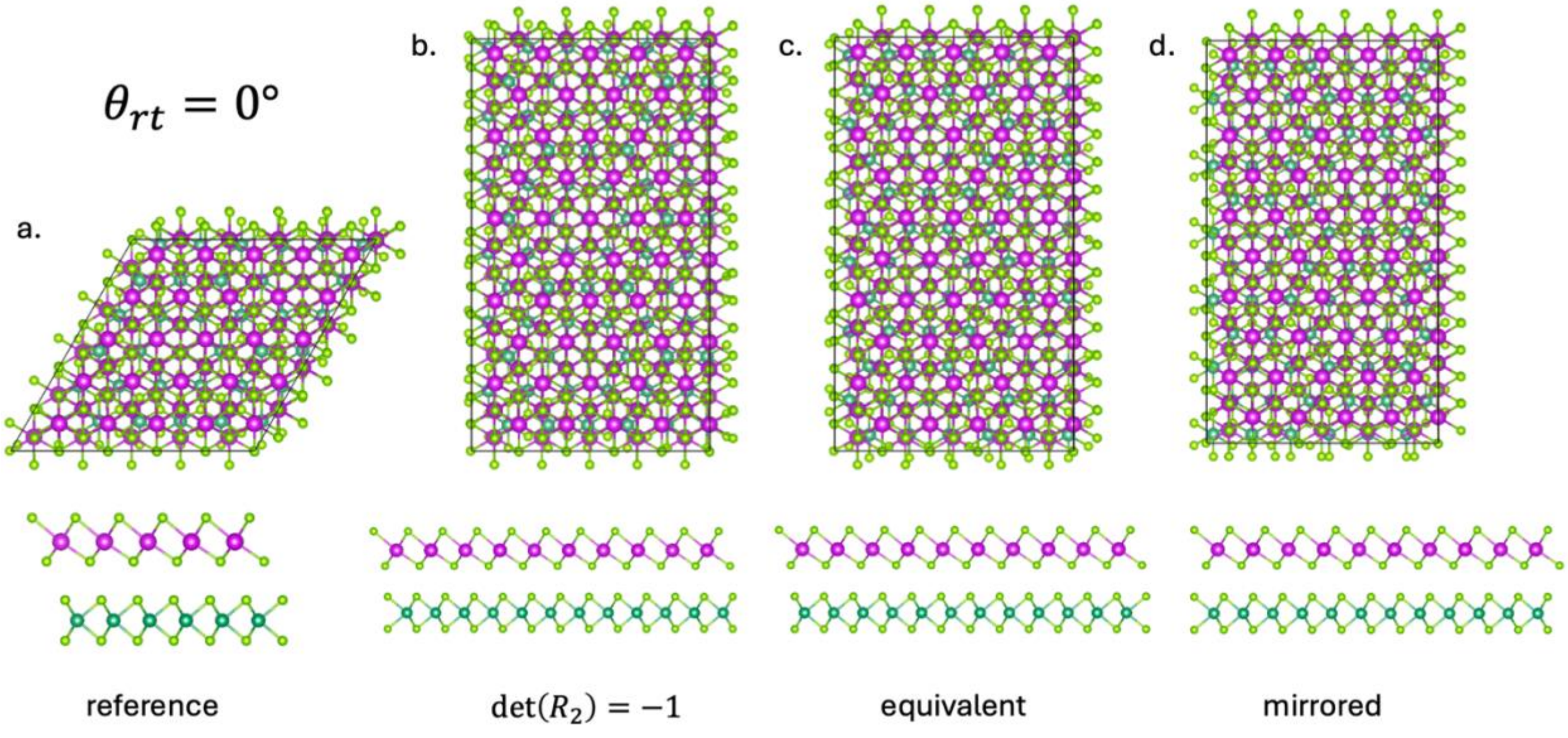

**Figure 3.** (a) The 466-atom heterostructure generated by lattice-match that represented the group $\theta_{rt} = 0°$. (b) A 932 atom supercell with a negative determinant for the rotation matrix, which was excluded from our workflow. (c) A 932 atom supercell equivalent to the 466 atom supercell. (d) A 932 atom supercell where the $NbSe_2$ monolayer at the interface mirrors the direction of the interfacial $NbSe_2$ monolayer in the reference. (a) and (c) are equivalent based on the structure-matching approach in pymatgen, as are (b) and (d). Structures are visualized with VESTA[55].

### c. RMG Input Parameters

Single-point calculations that include spin-orbit coupling (SOC) were performed using RMG's multigrid solver and *fully relativistic* norm-conserving PBE pseudopotentials (NC FR ONCVPSP v0.4) obtained from PseudoDojo. All non-SOC simulations (SCF, ion-optimized) used the Davidson algorithm and *non-relativistic* PBE pseudopotentials (SG15 ONCV) distributed with RMG. For RMG DFT calculations of 3D (e.g., $Bi_2Se_3$ and $NbSe_2$ bulk primitive cells) and 2D (2L-$Bi_2Se_3$/2L-$NbSe_2$ heterointerface supercells) structures, pyRMG was used to auto-generate wavefunction grids with a cutoff of 200 Rydberg and k-point meshes with a density of 0.2 $Bohr^{-1}$. We also included DFT-D3 dispersion corrections[56] to capture van der Waals interactions more correctly. For this investigation, we set the RMG energy convergence tolerance to $1\times10^{-10}$ Ha and the RMG RMS convergence tolerance to $1\times10^{-8}$ Ha.

We exclusively used the Davidson Kohn-Sham solver for high-throughput SCF calculations to ensure direct comparability of timings in this study (see Figure 4). Although the multigrid solver is expected to outperform Davidson for cells with tens of thousands of electrons, the exact crossover in performance is system and resource-dependent, as shown in Appendix B (Figure 13). We also disabled localized projectors for all calculations, though these can provide efficiency gains for larger cells. All calculations used Broyden charge-density mixing with a mixing factor of 0.1; this setting is more conservative than RMG's default of 0.5, which can increase runtimes but promotes convergence in difficult cells (e.g., 2D, large vacuum regions, anisotropy). Prior to ionic optimizations, atomic site positions were randomly displaced to break inherent symmetries, and lattices were fixed while ion positions were optimized.

# III. High-Throughput Workflow

## a. Workflow and Timing

After grouping the heterostructures by $\theta_{rt}$, the structures in each group with the fewest atoms were selected as representatives, resulting in 35 total structures for which we performed single-point (SCF) RMG DFT calculations. Figure 4 plots the SCF wall-clock time (x-axis) versus the true electrons-per-GPU value (y-axis) based on the heterostructure processor grids solved by pyRMG, where electrons_per_gpu=8 and --grid_divisibility_exponent=4 values were specified. For 32 of the 35 structures, pyRMG generates processor grids with true electrons-per-GPU values between ≈6.5 and 8.5, indicating that most values remain close to the target of 8. Points in Figure 4 are colored by total atom count, which in our dataset ranges from 168 to 1,046 atoms, and span a wide range of wall-clock times (from under 20 minutes to over 300 minutes).

The timings in this dataset are not strongly correlated with processor-grid anisotropy, $\mathcal{A}$, (Pearson $r \approx -0.322$) as our preliminary dataset (Appendix B), and are instead more correlated with the number of atoms in the system ($r \approx 0.705$). This is unsurprising, as the number of atoms determines the number of valence electrons and therefore the dimension of the projected subspace Hamiltonian that must be diagonalized. Calculations completed in under 5,000 s (blue region) contain at most 878 atoms, whereas those taking longer than 5,000 s (yellow region) contain at least 716 atoms. Timing profiles output by the base RMG code show that calculations in the yellow region are dominated by the diagonalization and matrix setup/reduce step subroutines of the Davidson solver, consistent with the scaling plot in the Appendix (Figure 15). For example, the diagonalization subroutine averages ≈67.3 s per electronic step (total runtime ≈15,290.7 s) for the 1,014-atom heterostructure, but only ≈13.2 s per electronic step (total runtime ≈3,600.8 s) for the 748-atom heterostructure.

Considering both the results presented here and those in Appendix B, we conclude that metrics such as the anisotropy, $\mathcal{A}$, and the number of atoms or total valence electrons can be useful to estimate the crossover point between Davidson and multigrid performance, but their sensitivity depends on the dataset. If a crossover point has been estimated (by $\mathcal{A}$, the number of atoms or another metric) and a few representative structures near this point have been identified, pyRMG can be used to automatically generate inputs, submit calculations, and measure timing differences across solvers (e.g., Davidson vs multigrid) or other input-file choices. By leveraging its automated input file generation and submission, users can rapidly test and tune their simulation inputs to maximize throughput before beginning production-scale RMG workflows.

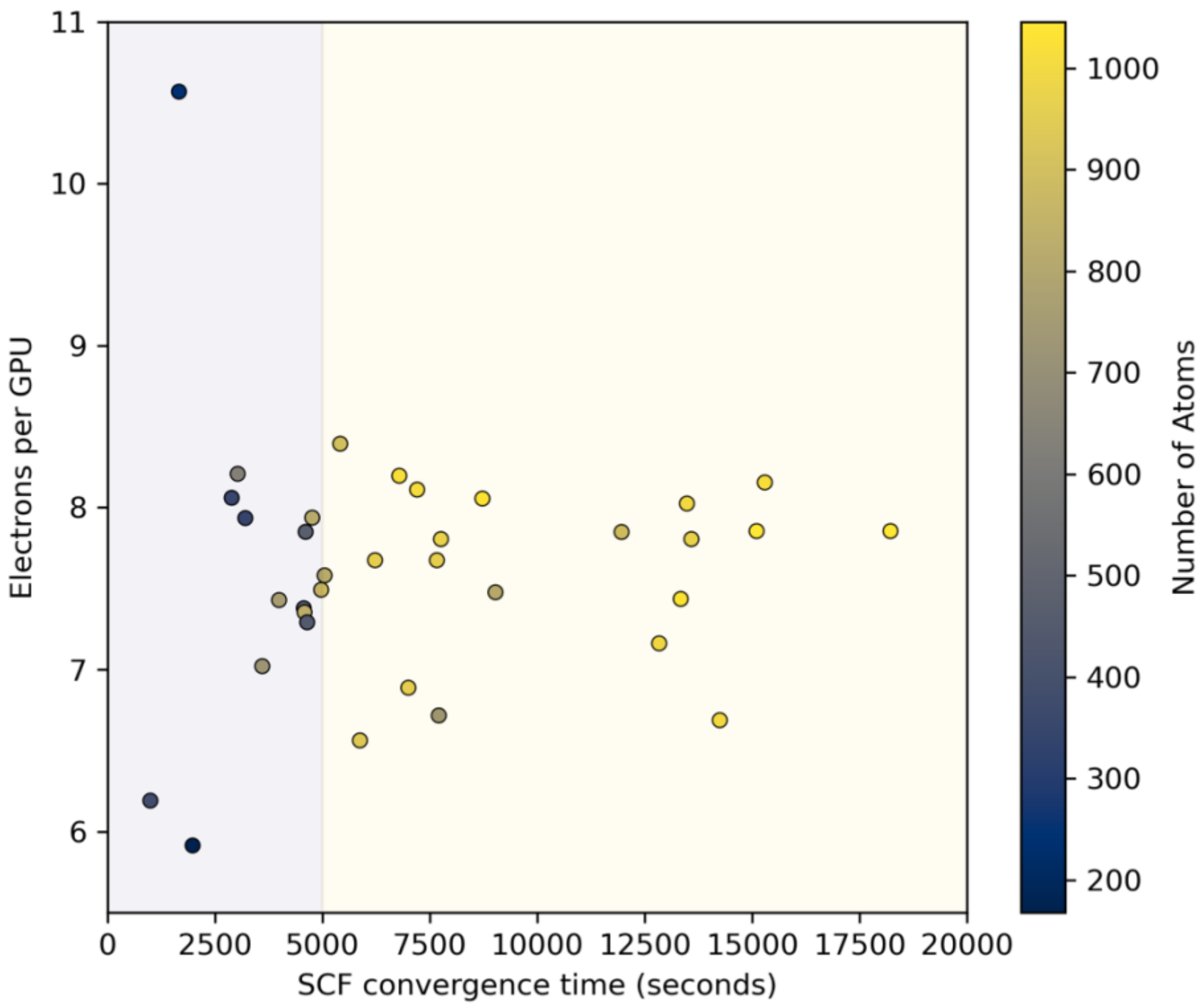

**Figure 4.** Single-point calculation convergence time vs. the electrons per GPU based on the processor grid solved by pyRMG, with color bars indicating (a) wavefunction grid anisotropy, $\mathcal{A}$, and (b) the number of atoms in the cell for the 35 2L-$Bi_2Se_3$/2L-$NbSe_2$ heterostructures generated via lattice-match. Calculation times are not as strongly correlated with $\mathcal{A}$ (r = -0.322) as they are with atom number (r = 0.705). We split the calculations into two regions based on the total number of atoms in the systems, represented by the blue and yellow overlaid boxes.

## b. Material Insights

*i. Single-Point Binding Energies, $\Delta E_{bind}$, versus $\varepsilon_{vM}$*

We first computed binding energies, $\Delta E_{bind}$, of all heterostructures relative to the isolated 2L-$NbSe_2$ and 2L-$Bi_2Se_3$ slabs from which they were constructed. This allows us to directly compare heterostructure stabilities, which *cannot* be done by simply normalizing their DFT energies on a per-atom or per-formula unit basis, because the lattice-matching procedure does not strictly preserve the $NbSe_2$/$Bi_2Se_3$ stoichiometry after primitive cells have been rescaled and matched according to lattice, angle and area tolerances. Binding energies were computed using

$$\Delta E_{bind} = \frac{E_{hetero} - xE_{2L-Bi_2Se_3} - yE_{2L-NbSe_2}}{xE_{2L-Bi_2Se_3}} \tag{17}$$

where $E_{hetero}$, $E_{2L-Bi_2Se_3}$ and $E_{2L-NbSe_2}$ are RMG DFT energies, and $x$ and $y$ are the number of primitive $2L - Bi_2Se_3$ and $2L - NbSe_2$ unit cells in the heterostructure. We chose to normalize $\Delta E_{bind}$ per $E_{2L-Bi_2Se_3}$, but we would expect similar energy trends if normalized to $E_{2L-NbSe_2}$.

Figure 5 compares binding energies, reduced twist angles, and von Mises strains, $\varepsilon_{vM}$, for our preliminary dataset (left, discussed in Appendix B) and our reported dataset (right). In the preliminary dataset, $\varepsilon_{vM}$ spans 0 to 0.069, and the least stable structure is destabilized by nearly 0.9 eV per 2L-$Bi_2Se_3$ relative to the most stable structure. In contrast, $\varepsilon_{vM}$ in our reported dataset spans only 0 to 0.016, with an energy variation of only 0.3 eV per 2L-$Bi_2Se_3$. The preliminary dataset exhibits a strong, positive correlation between $\Delta E_{bind}$ and $\varepsilon_{vM}$ (Pearson r ≈ 0.733), indicating that that larger strains correspond to more positive (less favorable) binding energies. This trend is not observed in our reported dataset, where the narrower strain range limits the predictive power of $\varepsilon_{vM}$. The von Mises strain is likely most sensitive when geometric distortions arising from lattice mismatch dominate the binding energy, which is not the case for experimentally relevant twist angles with small strains; this motivates evaluating twist energetics with first principles approaches such as RMG + pyRMG. As an aside, other studies have tried predicting strain-energy responses with elastic strain models[57], though we did not benchmark this material-specific metric here.

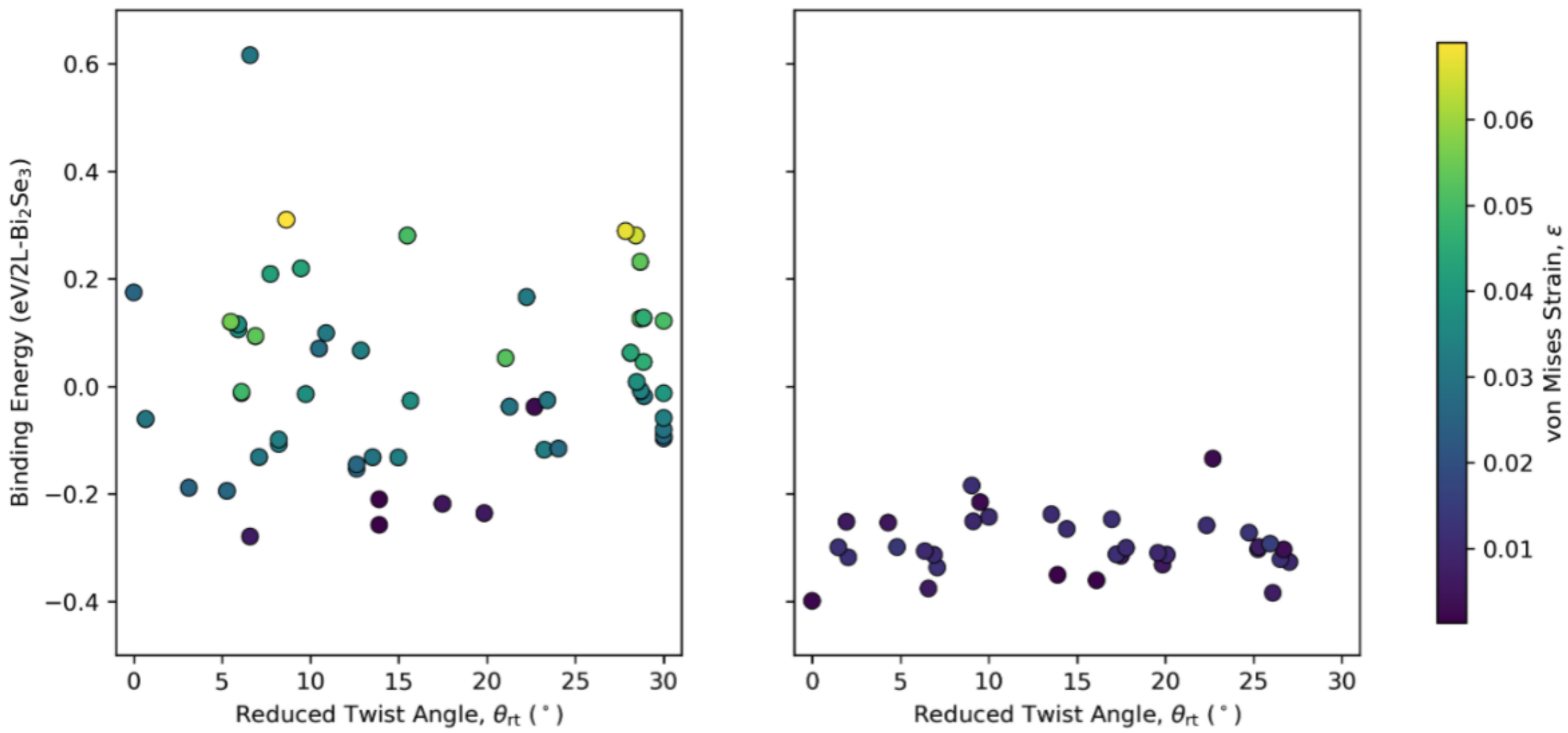

**Figure 5.** Reduced twist angle, $\theta_{rt}$ versus binding energy, $\Delta E_{bind}$, for (left) the 91 structures in our preliminary dataset spanning a large range of von Mises strains, and (right) the 35 structures in our reported dataset with a much narrower strain range. Binding energies are correlated with strain in the preliminary dataset (Pearson r = 0.733), but strain does not resolve energy differences within the reported dataset.

The three lowest-energy structures in our SCF dataset have reduced twist angles of $\theta_{rt}$ = 0° (466 atoms, reference structure in Figure 3), 26.1° (796 atoms), and 6.6° (352 atoms), with binding energies of -0.3990, -0.3841, and -0.3758 eV per 2L-$Bi_2Se_3$ unit cell, respectively. The 796-atom structure is an example of an unintuitively low-energy twisted heterostructure. It ranks 13th in both von Mises strain and fewest number of atoms in our reported dataset, yet it is only destabilized relative to the ground state heterostructure (untwisted, 0°) by 0.016 eV per 2L-

$Bi_2Se_3$. This system shows that large-cell 2L-$NbSe_2$/2L-$Bi_2Se_3$ heterostructures can express low energy twist angles that are as or more relevant than those expressed by smaller cells. Because low-energy twists may form during thin-film growth, it reinforces the importance of using linearly scaling *ab initio* methods that can handle large supercells. We also observe that certain small-twist structures, such as the 878-atom structure with $\theta_{rt}$ = 2.1° and the 862-atom structure with $\theta_{rt}$ = 1.5°, are only destabilized by 0.081 and 0.100 eV per 2L-$Bi_2Se_3$, respectively. For graphene homojunctions, small twist angles near 1° can exhibit "magic angle" properties[53], and thus one of the key advantages of pyRMG is its ability to simplify RMG calculation setup and submission for users exploring similarly large, computationally demanding structure models.

*ii. $\Delta E_{bind}$ Following Ionic Relaxation*

Ionic relaxation should reduce any large forces introduced by the geometric strain applied during lattice matching. To assess how relaxation affects the stability of heterostructures generated by our workflow, we fully relaxed the atomic structures and the layer-layer distances for six representative structures ($\theta_{rt}$ = 0.0°, 9.5°, 13.9°, 17.5°, 22.7°, 25.9°, with 466, 488, 168, 358, 244, and 656 atoms). For these systems, we find that the separation between the $NbSe_2$ and $Bi_2Se_3$ vdW-layers at the heterointerface changes by only about 0.038 Å, on average, likely because this distance was set based on the exfoliation energy curves (Appendix B). In contrast, interlayer separations within 2L-$NbSe_2$ and 2L-$Bi_2Se_3$ layers increase by 0.18 Å and 0.24 Å on average, clearly deviating from the bulk $NbSe_2$ and $Bi_2Se_3$ separations used to construct the heterostructures. This result shows that van der Waals gap distances away from the heterointerface should also be set using exfoliation energy curves.

Figure 7 compares the binding energies of the ion-optimized structures (black) to their unoptimized SCF counterparts (grey). As expected, all six heterostructures are stabilized upon relaxation, with an average stabilization of ≈0.119 eV per 2L-$Bi_2Se_3$. Although most structures exhibit similar stabilization magnitudes, some are larger than others; for example, the 244-atom structure ($\theta_{rt}$ = 22.7°) stabilizes by 0.142 eV per 2L-$Bi_2Se_3$, whereas the 466-atom structure ($\theta_{rt}$ = 0°) stabilizes by only 0.095 eV per 2L-$Bi_2Se_3$. Despite these shifts in absolute binding energies relative to the isolated bilayers, we observe no reordering in the relative stabilities of the twisted heterostructures: the most stable SCF structures remain the most stable after atomic relaxation. This preserved ordering suggests that (at least for the vdW-layered 2L-$NbSe_2$/2L-$Bi_2Se_3$ systems considered here, where no surface reconstruction is expected) single-point calculations provide a reasonable first approximation to vdW heterointerface energetics, without the computational expense of atomic relaxation.

*iii. Spin-Orbit Coupling, SOC*

All DFT results discussed thus far used non-relativistic pseudopotentials that neglect spin-orbit coupling (SOC). Because SOC captures the interaction between valence-electron spin–angular momentum and the effective magnetic field generated by core-electron motion, it is needed to reproduce band inversion in $Bi_2Se_3$ and related bulk tetradymite topological insulators, where Se p states that would normally form the valence-band maximum are pushed above Bi p states that would otherwise define the conduction-band minimum[58]. Although $Bi_2Se_3$ films grown on monolayer $NbSe_2$ do not exhibit the gapless Dirac surface states that arise from band inversion until their thickness exceeds three quintuple layers[47], and such calculations with two additional $Bi_2Se_3$ vdW layers would exceed our present computing allocation, Figure 6 shows that the electronic structure of 2L-$Bi_2Se_3$ is nonetheless strongly affected by SOC: its band gap shrinks from 0.69 eV to 0.17 eV.

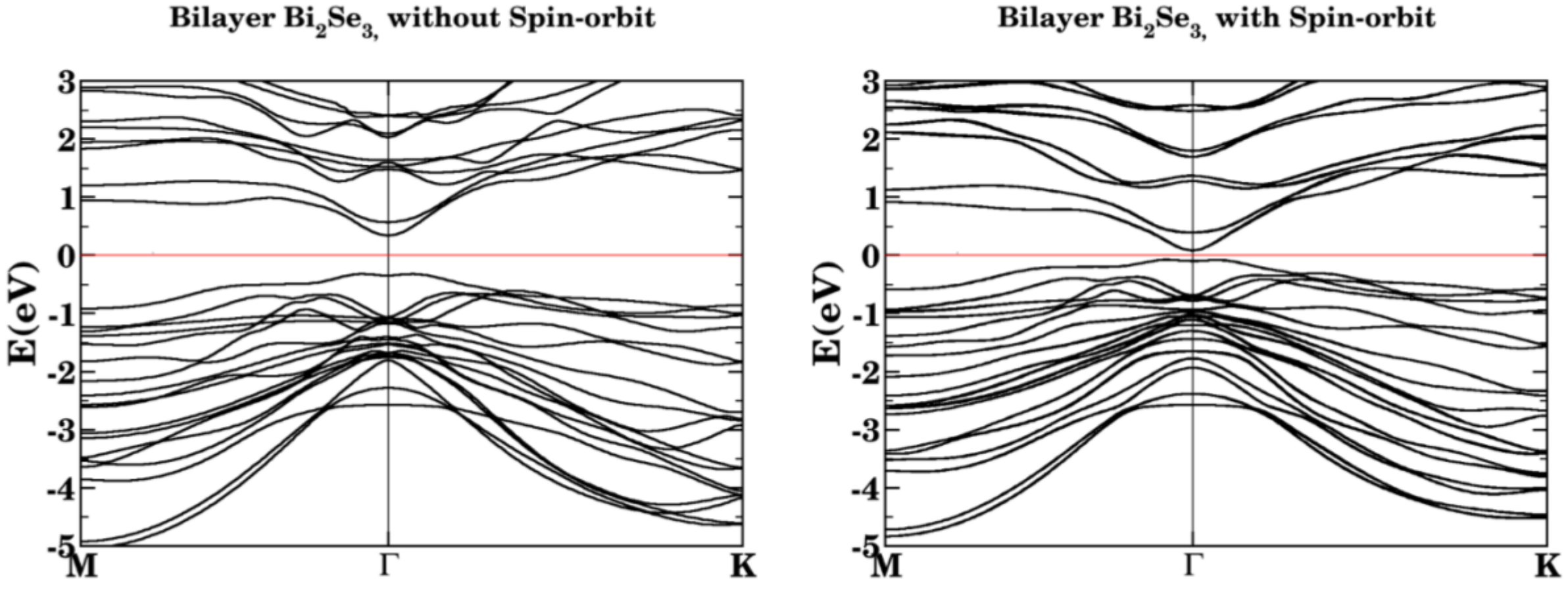

**Figure 6.** Band structure of 2L-$Bi_2Se_3$ along the M–Γ–K path, showing direct Γ-point band gaps (left) without SOC and (right) with SOC. Inclusion of SOC reduces the band gap from 0.69 eV to 0.17 eV.

Motivated by this, we performed SCF calculations using fully relativistic pseudopotentials for the six ion-relaxed $NbSe_2/Bi_2Se_3$ heterostructures. For large 2D supercells, however, SOC calculations are often prohibitive for high-throughput workflows, as they are typically an order of magnitude more expensive than PBE+D3 without SOC. For example, even when run on 5,376 GPUs, the 656-atom heterostructure required 40 SCF steps at ~532 s/step, totaling over 21,000 seconds. Nevertheless, it is notable that RMG can even converge SOC calculations of this scale, and pyRMG further streamlines their setup by automatically adjusting parameters such as the number of unoccupied electronic states to improve convergence. Using these capabilities, we successfully converged SOC SCF calculations for all six ion-optimized geometries.

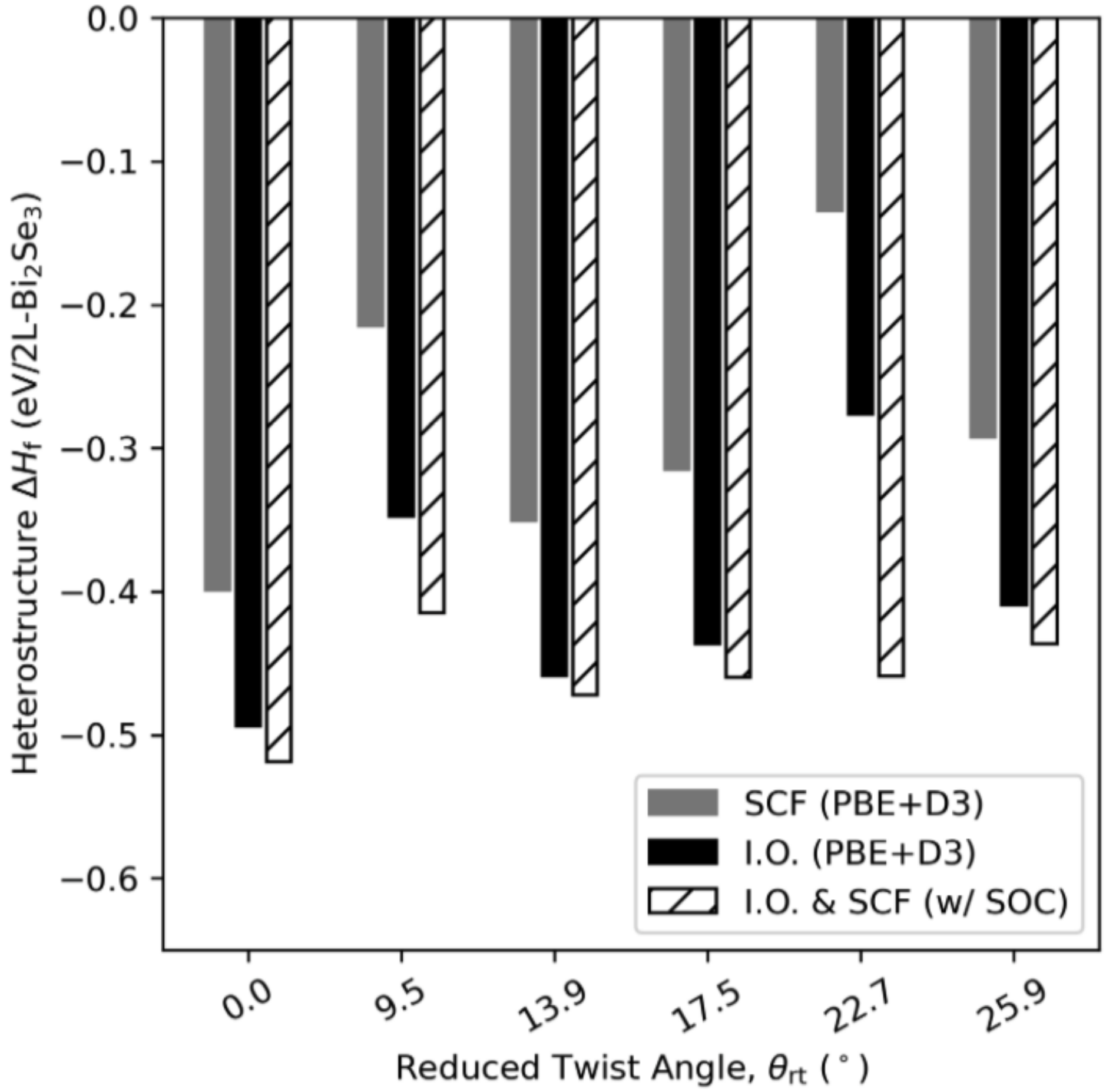

**Figure 7.** RMG DFT binding energies (normalized per 2L-$Bi_2Se_3$ unit) for self-consistent field (SCF), ion-optimized (I.O.), and SCF calculations including SOC, which were performed on the ion-optimized geometries.

Spin-orbit coupling stabilizes the binding energies of all six heterostructures relative to their non-SOC counterparts, though the magnitude varies with twist angle. The $\theta_{rt}$ = 22.7° structure is particularly stabilized, but this primarily arises from a change in k-point sampling: high-throughput single-point (SCF) and ion-optimized (I.O.) were performed at Γ, whereas the SOC calculation used a 2×2×1 mesh, substantially lowering the energy. This sensitivity illustrates a key challenge of variably sized supercells: even in high-throughput studies, k-point meshes must be benchmarked carefully, as inappropriate densities can result in poorly converged energetics. For systems where k-point meshes were fixed, the relative ordering energetics remains unchanged across SCF, ion-optimized, and SOC-inclusive calculations. We do not expect this trend to be universal, as these six structures were selected because they spanned the full range of SCF energies. For our broad energy window, however, we observe that SCF captures the same stability trends as more expensive atom-relaxed and SOC calculations, making them a useful first-pass screening tool.

As demonstrated by the position-resolved DOS for the 244-atom structure with SOC (Figure 8), 2L-$NbSe_2$ retains metallic states at the Fermi level, while 2L-$Bi_2Se_3$ exhibits a finite gap consistent with Figure 6. The 244-atom heterostructure and all others in our datasets should therefore be metallic. Γ-only calculations can converge the energies of large supercells (Figure 7), while the insufficiently dense k-point meshes can nonetheless miss essential features of the primitive-cell electron structure, including Nb-based metallic states crossing at the Fermi level. Additionally, the position-resolved DOS reveals 2L-$Bi_2Se_3$–derived states near the Fermi level at the heterointerface. This suggests that electronic overlap between the layers may facilitate charge transfer, which we analyze in the following section.

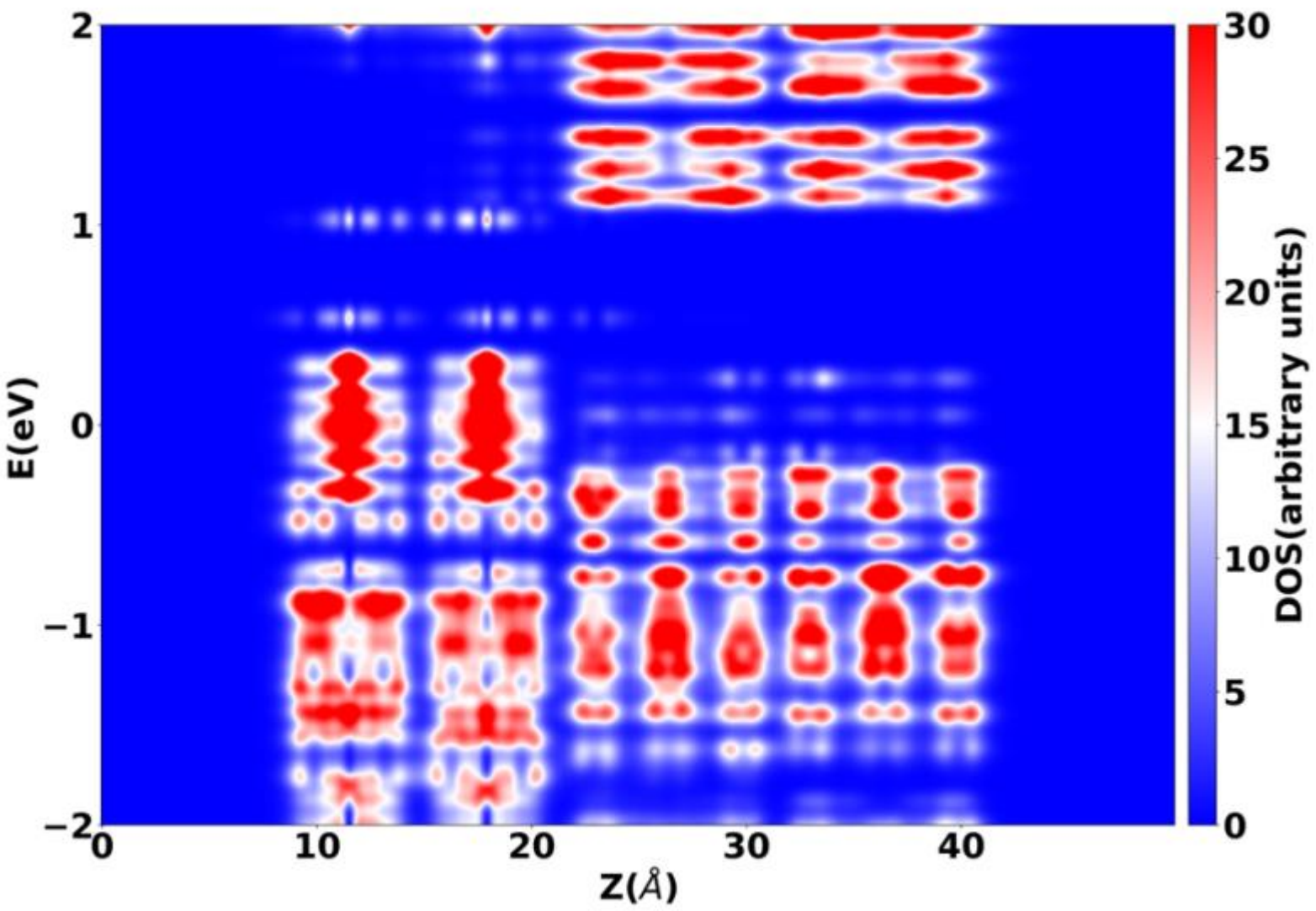

**Figure 8.** Position-resolved density of states for the 244-atom heterostructure ($\theta_{rt}$ = 22.7°). The 2L-$NbSe_2$ region lies at Z < 21Å, and the 2L-$Bi_2Se_3$ region at Z > 21Å. The 2L-$NbSe_2$ layers exhibit high concentrations of electronic states at the Fermi level (0.0 eV), consistent with metallic character.

*iv. Charge Transfer at the Heterointerface*

Charge transfer at $Bi_2Se_3$ heterointerfaces can result from band realignment that produces Schottky barriers[59], which in turn can lead to work-function shifts[60], altered carrier mobilities[61], and interfacial dipoles that influence topological properties[62]. In metallic/semiconducting stacks, such interlayer charge redistribution could modify the electronic structure near the Fermi level, which might affect the relative stability of different twists or registries. Because 2D $NbSe_2$ is metallic[63], while our DFT results show that bilayer $Bi_2Se_3$ is a small-gap semiconductor, some interfacial charge transfer is expected. However, the magnitude and direction of this charge transfer, whether it correlates with energetic stability, varies between SCF and atom-relaxed geometries, or is influenced by SOC is not immediately clear.

Figure 9 compares the net charge per vdW layer across the six twisted heterostructures for all three calculation types. Layer-resolved charges were obtained by summing Voronoi-based on-site charges partitioned with the base RMG code, and then normalizing these summed charges by the number of primitive unit cells contained in each layer (e.g., the 168-atom heterostructure contains 13 and 9 primitive cells for its 2L-$NbSe_2$ and 2L-$Bi_2Se_3$ components, respectively). For SCF and ion-optimized calculations without SOC, the net layer charges follow nearly identical trends, and in many cases the values are almost indistinguishable. $NbSe_2$ layers consistently carry a net negative charge, while $Bi_2Se_3$ layers carry a net positive charge, indicating that metallic $NbSe_2$ accepts electron density from $Bi_2Se_3$. Although no correlations between layer charge and interfacial energetics emerge, the SOC-inclusive calculations performed on ion-optimized geometries display substantially larger variation in the magnitude of charge transfer, particularly for vdW layers at the interface.

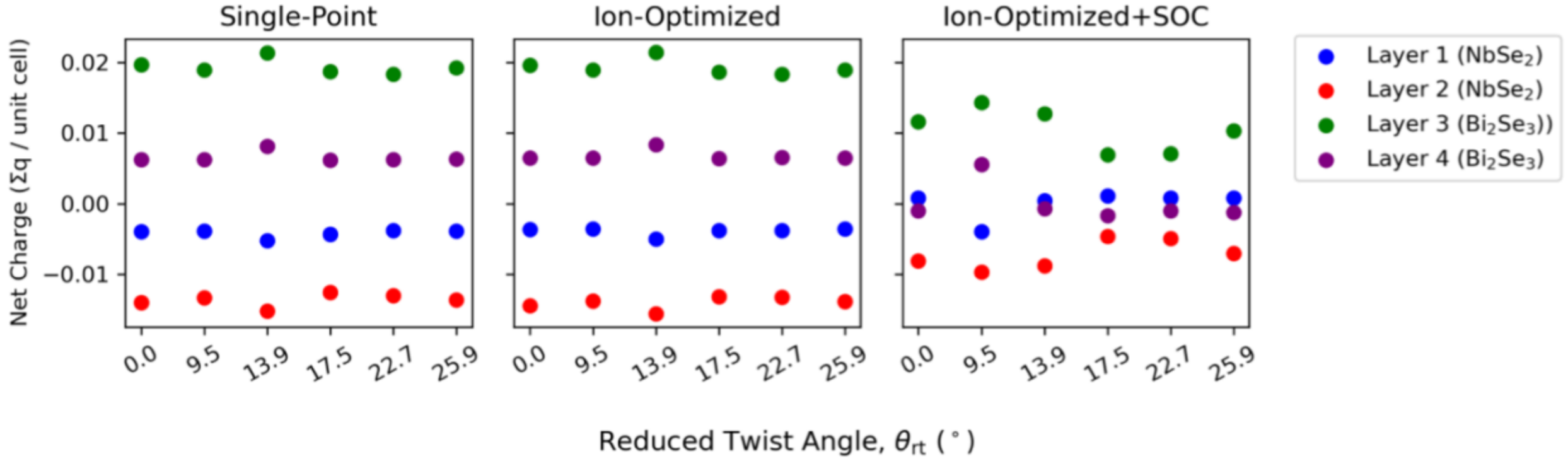

**Figure 9.** Net charge (normalized per unit cell) for the vdW layers in the six heterostructures considered. For all structures, vdW monolayers at the heterointerface have larger net charges than monolayers away from the interface, suggesting some charge transfer. Differences in layer charges between the different twist angles are much more pronounced for calculations including SOC.

# Discussion

In real epitaxial growth, films (e.g., $Bi_2Se_3$) are often grown on substrates (e.g., $NbSe_2$) that have been deposited on underlying crystalline materials. Layers in these stacks impose epitaxial strain on adjacent layers via lattice mismatch, which can propagate through the heterostructure. Simulating strain effects in 2L-$NbSe_2$/2L-$Bi_2Se_3$ interfaces, which might be compounded or off-set based on twist, for example, represents a straightforward extension of our current investigation. HeteroBuilder allows users to build multi-layer, lattice-matched stacks, and pyRMG can then be used to broadcast input generation, integrate with a job scheduler and submit these calculations on leadership computing platforms. Rather than fixing all calculations to the Davidson solver (as done here for benchmarking purposes), future studies could make broader use of RMG's weakly linearly-scaling multigrid algorithm, which offers improved convergence for large or variable-sized supercells.

Beyond 2D heterointerfaces, the scalability of the RMG + pyRMG workflow makes it well-suited to supplement existing computational materials databases, particularly with important but underrepresented materials systems. As noted earlier, these might include defect systems (point defects, offset interfaces, grain boundaries), off-stoichiometric phases, or high-entropy alloys. More broadly, it can enable DFT calculations of snapshots taken from large-scale classical molecular dynamics simulations, which routinely involve thousands of atoms. These calculations, which would be otherwise inaccessible to plane-wave based DFT codes could be used to better validate data-driven methods, such as machine-learned force fields. In short, by better leveraging exascale computing resources, RMG + pyRMG helps bridge the gap between first-principles accuracy and large system modeling.

## IV. Summary and Conclusions

We developed pyRMG to automate and scale RMG DFT calculations on leadership-class CPU-GPU architectures. The package streamlines input generation, improves restart and resubmission behavior, and automatically constructs processor grids that yield more homogeneous subdomains and faster runtimes. It also gives users explicit control over resource allocation by automating processor grid construction based on electrons-per-GPU targets and grid-divisibility constraints, and integrates with adaptive schedulers like MatEnsemble to manage RMG workflows across scales. By handling common RMG failures, such as insufficient memory, poor grid choices, and grid incompatibilities during relaxation, pyRMG reduces user overhead and enhances the reliability of high-throughput campaigns.

We benchmarked pyRMG on a challenging study of 2L-$Bi_2Se_3$/2L-$NbSe_2$ heterostructures, characterized by their large vacuum regions, extreme anisotropy, and widely varying valence electron counts. Using pyRMG with a single YAML configuration file, RMG achieved SCF convergence for all 35 structures. Analysis of these structures reveals several insights: strain correlates with energetics but shows minimal predictive power at smaller scales; ion relaxation and spin-orbit coupling stabilize the heterostructures relative to isolated bilayers; and the position-dependent DOS indicates $Bi_2Se_3$ states near the Fermi level. This results in negative charge transfer at the heterointerface from $Bi_2Se_3$ to $NbSe_2$ monolayers that is most sensitive to twist angle when SOC is included. Overall, pyRMG allows users to harness RMG's strong-scaling behavior, making high-throughput materials discovery workflows more practical on HPC systems.

## V. Acknowledgments

This work is part of the Quantum Correlated Materials Automated Discovery (QCAD) project supported by the U.S. Department of Energy (DOE) contract DE-AC05-00OR22725. QCAD is part of the INTERSECT initiation under the Laboratory Directed Research and Development (LDRD) program at Oak Ridge National Laboratory (ORNL), which is managed by UT-Battelle, LLC. This research used the *Frontier* supercomputer, which is managed by the Oak Ridge Leadership Computing Facility (OLCF) at ORNL.

## VI. Code and Data Availability

The **Q-CAD** platform (https://github.com/Q-CAD) contains the **HeteroBuilder** Python package (https://github.com/Q-CAD/HeteroBuilder) used to generate heterostructure structures for this work. The structure and RMG output files are available in the 'data' submodule of

https://github.com/Q-CAD/pyRMG, which includes a Jupyter notebook to parse and visualize the dataset and generate the figures presented in the main text and Supplementary Appendix B. The Jupyter notebook requires the **parse2fit** Python package to parse RMG outputs, which is also available in the **Q-CAD** repository.

## VII. Author Contributions

RM developed **pyRMG** and **HeteroBuilder**, generated the datasets, performed the high-throughput heterostructure calculations, analyzed and visualized the data, and drafted the manuscript. SB developed **MatEnsemble** and provided expertise on adaptive scheduling and high-performance computing best practices for high-throughput workflows. EB, WL and JB developed and updated RMG to converge this workflow. EB performed the Davidson vs. multigrid algorithm scaling analysis and SOC calculations. WL created the 2L-$Bi_2Se_3$ band structure plots and position-resolved DOS for 2L-$Bi_2Se_3$/2L-$NbSe_2$. JB contributed to manuscript writing. PG conceived of and supervised the project and contributed to manuscript writing.

## VIII. References

[1] A. Jain, S.P. Ong, G. Hautier, W. Chen, W.D. Richards, S. Dacek, S. Cholia, D. Gunter, D. Skinner, G. Ceder, and K.A. Persson, “Commentary: The Materials Project: A materials genome approach to accelerating materials innovation,” APL Mater. **1**(1), 011002 (2013).
[2] K. Choudhary, K.F. Garrity, A.C.E. Reid, B. DeCost, A.J. Biacchi, A.R. Hight Walker, Z. Trautt, J. Hattrick-Simpers, A.G. Kusne, A. Centrone, A. Davydov, J. Jiang, R. Pachter, G. Cheon, E. Reed, A. Agrawal, X. Qian, V. Sharma, H. Zhuang, S.V. Kalinin, B.G. Sumpter, G. Pilania, P. Acar, S. Mandal, K. Haule, D. Vanderbilt, K. Rabe, and F. Tavazza, “The joint automated repository for various integrated simulations (JARVIS) for data-driven materials design,” Npj Comput. Mater. **6**(1), 173 (2020).
[3] S. Kirklin, J.E. Saal, B. Meredig, A. Thompson, J.W. Doak, M. Aykol, S. Rühl, and C. Wolverton, “The Open Quantum Materials Database (OQMD): assessing the accuracy of DFT formation energies,” Npj Comput. Mater. **1**(1), 15010 (2015).
[4] L. Sbailò, Á. Fekete, L.M. Ghiringhelli, and M. Scheffler, “The NOMAD Artificial-Intelligence Toolkit: turning materials-science data into knowledge and understanding,” Npj Comput. Mater. **8**(1), 250 (2022).

[5] S. Curtarolo, W. Setyawan, S. Wang, J. Xue, K. Yang, R.H. Taylor, L.J. Nelson, G.L.W. Hart, S. Sanvito, M. Buongiorno-Nardelli, N. Mingo, and O. Levy, "AFLOWLIB.ORG: A distributed materials properties repository from high-throughput *ab initio* calculations," Comput. Mater. Sci. **58**, 227–235 (2012).
[6] V. Stevanović, S. Lany, X. Zhang, and A. Zunger, "Correcting density functional theory for accurate predictions of compound enthalpies of formation: Fitted elemental-phase reference energies," Phys. Rev. B **85**(11), 115104 (2012).
[7] Z.J.L. Bare, R.J. Morelock, and C.B. Musgrave, "Dataset of theoretical multinary perovskite oxides," Sci. Data **10**(1), 244 (2023).
[8] Y.G. Chung, J. Camp, M. Haranczyk, B.J. Sikora, W. Bury, V. Krungleviciute, T. Yildirim, O.K. Farha, D.S. Sholl, and R.Q. Snurr, "Computation-Ready, Experimental Metal–Organic Frameworks: A Tool To Enable High-Throughput Screening of Nanoporous Crystals," Chem. Mater. **26**(21), 6185–6192 (2014).
[9] G. Hautier, A. Jain, H. Chen, C. Moore, S.P. Ong, and G. Ceder, "Novel mixed polyanions lithium-ion battery cathode materials predicted by high-throughput ab initio computations," J. Mater. Chem. **21**(43), 17147–17153 (2011).
[10] K.T. Winther, M.J. Hoffmann, J.R. Boes, O. Mamun, M. Bajdich, and T. Bligaard, "Catalysis-Hub.org, an open electronic structure database for surface reactions," Sci. Data **6**(1), 75 (2019).
[11] L. Ward, A. Dunn, A. Faghaninia, N.E.R. Zimmermann, S. Bajaj, Q. Wang, J. Montoya, J. Chen, K. Bystrom, M. Dylla, K. Chard, M. Asta, K.A. Persson, G.J. Snyder, I. Foster, and A. Jain, "Matminer: An open source toolkit for materials data mining," Comput. Mater. Sci. **152**, 60–69 (2018).
[12] A. Jain, G. Hautier, S.P. Ong, and K. Persson, "New opportunities for materials informatics: Resources and data mining techniques for uncovering hidden relationships," J. Mater. Res. **31**(8), 977–994 (2016).
[13] C.J. Bartel, S.L. Millican, A.M. Deml, J.R. Rumptz, W. Tumas, A.W. Weimer, S. Lany, V. Stevanović, C.B. Musgrave, and A.M. Holder, "Physical descriptor for the Gibbs energy of inorganic crystalline solids and temperature-dependent materials chemistry," Nat. Commun. **9**(1), 4168 (2018).
[14] M. Aykol, S.S. Dwaraknath, W. Sun, and K.A. Persson, "Thermodynamic limit for synthesis of metastable inorganic materials," Sci. Adv. **4**(4), eaaq0148 (2018).
[15] S. Lany, "Band-structure calculations for the 3d transition metal oxides in GW," Phys. Rev. B **87**(8), 085112 (2013).
[16] K. Choudhary, Q. Zhang, A.C.E. Reid, S. Chowdhury, N. Van Nguyen, Z. Trautt, M.W. Newrock, F.Y. Congo, and F. Tavazza, "Computational screening of high-performance optoelectronic materials using OptB88vdW and TB-mBJ formalisms," Sci. Data **5**, 180082 (2018).
[17] M.J. McDermott, S.S. Dwaraknath, and K.A. Persson, "A graph-based network for predicting chemical reaction pathways in solid-state materials synthesis," Nat. Commun. **12**(1), 3097 (2021).
[18] N.J. Szymanski, B. Rendy, Y. Fei, R.E. Kumar, T. He, D. Milsted, M.J. McDermott, M. Gallant, E.D. Cubuk, A. Merchant, H. Kim, A. Jain, C.J. Bartel, K. Persson, Y. Zeng, and G. Ceder, "An autonomous laboratory for the accelerated synthesis of novel materials," Nature **624**(7990), 86–91 (2023).
[19] O. Mamun, K.T. Winther, J.R. Boes, and T. Bligaard, "High-throughput calculations of catalytic properties of bimetallic alloy surfaces," Sci. Data **6**(1), 76 (2019).

[20] I. Batatia, D.P. Kovács, G.N.C. Simm, C. Ortner, and G. Csányi, "MACE: Higher Order Equivariant Message Passing Neural Networks for Fast and Accurate Force Fields," (n.d.).
[21] B. Deng, P. Zhong, K. Jun, J. Riebesell, K. Han, C.J. Bartel, and G. Ceder, "CHGNet as a pretrained universal neural network potential for charge-informed atomistic modelling," Nat. Mach. Intell. **5**(9), 1031–1041 (2023).
[22] A. Musaelian, S. Batzner, A. Johansson, L. Sun, C.J. Owen, M. Kornbluth, and B. Kozinsky, "Learning local equivariant representations for large-scale atomistic dynamics," Nat. Commun. **14**(1), 579 (2023).
[23] S. Batzner, A. Musaelian, L. Sun, M. Geiger, J.P. Mailoa, M. Kornbluth, N. Molinari, T.E. Smidt, and B. Kozinsky, "E(3)-equivariant graph neural networks for data-efficient and accurate interatomic potentials," Nat. Commun. **13**(1), 2453 (2022).
[24] J. Behler, "Perspective: Machine learning potentials for atomistic simulations," J. Chem. Phys. **145**(17), 170901 (2016).
[25] J. Riebesell, R.E.A. Goodall, P. Benner, Y. Chiang, B. Deng, G. Ceder, M. Asta, A.A. Lee, A. Jain, and K.A. Persson, "Matbench Discovery -- A framework to evaluate machine learning crystal stability predictions," (2024).
[26] X.-G. Zhao, G.M. Dalpian, Z. Wang, and A. Zunger, "Polymorphous nature of cubic halide perovskites," Phys. Rev. B **101**(15), 155137 (2020).
[27] O.I. Malyi, and A. Zunger, "False metals, real insulators, and degenerate gapped metals," Appl. Phys. Rev. **7**(4), 041310 (2020).
[28] Z. Zhao, B. Austin, S. Maintz, and M. Marsman, "VASP Performance on HPE Cray EX Based on NVIDIA A100 GPUs and AMD Milan CPUs," (n.d.).
[29] E.L. Briggs, D.J. Sullivan, and J. Bernholc, "Real-space multigrid-based approach to large-scale electronic structure calculations," Phys. Rev. B **54**(20), 14362–14375 (1996).
[30] E.L. Briggs, W. Lu, and J. Bernholc, "Adaptive finite differencing in high accuracy electronic structure calculations," Npj Comput. Mater. **10**(1), 1–9 (2024).
[31] K. Lejaeghere, G. Bihlmayer, T. Björkman, P. Blaha, S. Blügel, V. Blum, D. Caliste, I.E. Castelli, S.J. Clark, A.D. Corso, S. de Gironcoli, T. Deutsch, J.K. Dewhurst, I.D. Marco, C. Draxl, M. Dułak, O. Eriksson, J.A. Flores-Livas, K.F. Garrity, L. Genovese, P. Giannozzi, M. Giantomassi, S. Goedecker, X. Gonze, O. Grånäs, E.K.U. Gross, A. Gulans, F. Gygi, D.R. Hamann, P.J. Hasnip, N. a. W. Holzwarth, D. Iuşan, D.B. Jochym, F. Jollet, D. Jones, G. Kresse, K. Koepernik, E. Küçükbenli, Y.O. Kvashnin, I.L.M. Locht, S. Lubeck, M. Marsman, N. Marzari, U. Nitzsche, L. Nordström, T. Ozaki, L. Paulatto, C.J. Pickard, W. Poelmans, M.I.J. Probert, K. Refson, M. Richter, G.-M. Rignanese, S. Saha, M. Scheffler, M. Schlipf, K. Schwarz, S. Sharma, F. Tavazza, P. Thunström, A. Tkatchenko, M. Torrent, D. Vanderbilt, M.J. van Setten, V.V. Speybroeck, J.M. Wills, J.R. Yates, G.-X. Zhang, and S. Cottenier, "Reproducibility in density functional theory calculations of solids," Science **351**(6280), aad3000 (2016).
[32] J. Kim, A.D. Baczewski, T.D. Beaudet, A. Benali, M.C. Bennett, M.A. Berrill, N.S. Blunt, E.J.L. Borda, M. Casula, D.M. Ceperley, S. Chiesa, B.K. Clark, R.C. Clay, K.T. Delaney, M. Dewing, K.P. Esler, H. Hao, O. Heinonen, P.R.C. Kent, J.T. Krogel, I. Kylänpää, Y.W. Li, M.G. Lopez, Y. Luo, F.D. Malone, R.M. Martin, A. Mathuriya, J. McMinis, C.A. Melton, L. Mitas, M.A. Morales, E. Neuscamman, W.D. Parker, S.D. Pineda Flores, N.A. Romero, B.M. Rubenstein, J.A.R. Shea, H. Shin, L. Shulenburger, A.F. Tillack, J.P. Townsend, N.M. Tubman, B. Van Der Goetz, J.E. Vincent, D.C. Yang, Y. Yang, S. Zhang, and L. Zhao, "QMCPACK: an open source ab initio quantum Monte Carlo package for the electronic structure of atoms, molecules and solids," J. Phys. Condens. Matter **30**(19), 195901 (2018).

[33] P.R.C. Kent, A. Annaberdiyev, A. Benali, M.C. Bennett, E.J. Landinez Borda, P. Doak, H. Hao, K.D. Jordan, J.T. Krogel, I. Kylänpää, J. Lee, Y. Luo, F.D. Malone, C.A. Melton, L. Mitas, M.A. Morales, E. Neuscamman, F.A. Reboredo, B. Rubenstein, K. Saritas, S. Upadhyay, G. Wang, S. Zhang, and L. Zhao, "QMCPACK: Advances in the development, efficiency, and application of auxiliary field and real-space variational and diffusion quantum Monte Carlo," J. Chem. Phys. **152**(17), 174105 (2020).
[34] S.P. Ong, W.D. Richards, A. Jain, G. Hautier, M. Kocher, S. Cholia, D. Gunter, V.L. Chevrier, K.A. Persson, and G. Ceder, "Python Materials Genomics (pymatgen): A robust, open-source python library for materials analysis," Comput. Mater. Sci. **68**, 314–319 (2013).
[35] A. Hjorth Larsen, J. Jørgen Mortensen, J. Blomqvist, I.E. Castelli, R. Christensen, M. Dułak, J. Friis, M.N. Groves, B. Hammer, C. Hargus, E.D. Hermes, P.C. Jennings, P. Bjerre Jensen, J. Kermode, J.R. Kitchin, E. Leonhard Kolsbjerg, J. Kubal, K. Kaasbjerg, S. Lysgaard, J. Bergmann Maronsson, T. Maxson, T. Olsen, L. Pastewka, A. Peterson, C. Rostgaard, J. Schiøtz, O. Schütt, M. Strange, K.S. Thygesen, T. Vegge, L. Vilhelmsen, M. Walter, Z. Zeng, and K.W. Jacobsen, "The atomic simulation environment—a Python library for working with atoms," J. Phys. Condens. Matter **29**(27), 273002 (2017).
[36] G. Kresse, and J. Hafner, "Ab initio molecular dynamics for liquid metals," Phys. Rev. B **47**(1), 558–561 (1993).
[37] G. Kresse, and J. Furthmüller, "Efficiency of ab-initio total energy calculations for metals and semiconductors using a plane-wave basis set," Comput. Mater. Sci. **6**(1), 15–50 (1996).
[38] G. Kresse, and J. Furthmüller, "Efficient iterative schemes for ab initio total-energy calculations using a plane-wave basis set," Phys. Rev. B **54**(16), 11169–11186 (1996).
[39] P. Giannozzi, S. Baroni, N. Bonini, M. Calandra, R. Car, C. Cavazzoni, D. Ceresoli, G.L. Chiarotti, M. Cococcioni, I. Dabo, A. Dal Corso, S. de Gironcoli, S. Fabris, G. Fratesi, R. Gebauer, U. Gerstmann, C. Gougoussis, A. Kokalj, M. Lazzeri, L. Martin-Samos, N. Marzari, F. Mauri, R. Mazzarello, S. Paolini, A. Pasquarello, L. Paulatto, C. Sbraccia, S. Scandolo, G. Sclauzero, A.P. Seitsonen, A. Smogunov, P. Umari, and R.M. Wentzcovitch, "QUANTUM ESPRESSO: a modular and open-source software project for quantum simulations of materials," J. Phys. Condens. Matter Inst. Phys. J. **21**(39), 395502 (2009).
[40] P. Giannozzi, O. Andreussi, T. Brumme, O. Bunau, M.B. Nardelli, M. Calandra, R. Car, C. Cavazzoni, D. Ceresoli, M. Cococcioni, N. Colonna, I. Carnimeo, A.D. Corso, S.D. Gironcoli, P. Delugas, R.A. Distasio, A. Ferretti, A. Floris, G. Fratesi, G. Fugallo, R. Gebauer, U. Gerstmann, F. Giustino, T. Gorni, J. Jia, M. Kawamura, H.Y. Ko, A. Kokalj, E. Kücükbenli, M. Lazzeri, M. Marsili, N. Marzari, F. Mauri, N.L. Nguyen, H.V. Nguyen, A. Otero-De-La-Roza, L. Paulatto, S. Poncé, D. Rocca, R. Sabatini, B. Santra, M. Schlipf, A.P. Seitsonen, A. Smogunov, I. Timrov, T. Thonhauser, P. Umari, N. Vast, X. Wu, and S. Baroni, "Advanced capabilities for materials modelling with Quantum ESPRESSO," J. Phys. Condens. Matter **29**(46), 465901 (2017).
[41] S. Bagchi, A. Biswas, P.V. Balachandran, A. Ghosh, and P. Ganesh, "Towards 'on-demand' van der Waals epitaxy with hpc-driven online ensemble sampling," (2025).
[42] A.M. Ganose, H. Sahasrabuddhe, M. Asta, K. Beck, T. Biswas, A. Bonkowski, J. Bustamante, X. Chen, Y. Chiang, D.C. Chrzan, J. Clary, O.A. Cohen, C. Ertural, M.C. Gallant, J. George, S. Gerits, R.E.A. Goodall, R.D. Guha, G. Hautier, M. Horton, T.J. Inizan, A.D. Kaplan, R.S. Kingsbury, M.C. Kuner, B. Li, X. Linn, M.J. McDermott, R.S. Mohanakrishnan, A.N. Naik, J.B. Neaton, S.M. Parmar, K.A. Persson, G. Petretto, T.A.R. Purcell, F. Ricci, B. Rich, J. Riebesell, G.-M. Rignanese, A.S. Rosen, M. Scheffler, J. Schmidt, J.-X. Shen, A. Sobolev, R. Sundararaman, C. Tezak, V. Trinquet, J.B. Varley, D. Vigil-Fowler, D. Wang, D. Waroquiers,

M. Wen, H. Yang, H. Zheng, J. Zheng, Z. Zhu, and A. Jain, "Atomate2: modular workflows for materials science," Digit. Discov. **4**(7), 1944–1973 (2025).
[43] K. Mathew, J.H. Montoya, A. Faghaninia, S. Dwarakanath, M. Aykol, H. Tang, I. Chu, T. Smidt, B. Bocklund, M. Horton, J. Dagdelen, B. Wood, Z.-K. Liu, J. Neaton, S.P. Ong, K. Persson, and A. Jain, "Atomate: A high-level interface to generate, execute, and analyze computational materials science workflows," Comput. Mater. Sci. **139**, 140–152 (2017).
[44] K. Mazumder, and P.M. Shirage, "A brief review of Bi2Se3 based topological insulator: From fundamentals to applications," J. Alloys Compd. **888**, 161492 (2021).
[45] X. Xi, Z. Wang, W. Zhao, J.-H. Park, K.T. Law, H. Berger, L. Forró, J. Shan, and K.F. Mak, "Ising pairing in superconducting NbSe2 atomic layers," Nat. Phys. **12**(2), 139–143 (2016).
[46] W. Dai, A. Richardella, R. Du, W. Zhao, X. Liu, C.X. Liu, S.-H. Huang, R. Sankar, F. Chou, N. Samarth, and Q. Li, "Proximity-effect-induced Superconducting Gap in Topological Surface States – A Point Contact Spectroscopy Study of NbSe2/Bi2Se3 Superconductor-Topological Insulator Heterostructures," Sci. Rep. **7**(1), 7631 (2017).
[47] H. Yi, L.-H. Hu, Y. Wang, R. Xiao, J. Cai, D.R. Hickey, C. Dong, Y.-F. Zhao, L.-J. Zhou, R. Zhang, A.R. Richardella, N. Alem, J.A. Robinson, M.H.W. Chan, X. Xu, N. Samarth, C.-X. Liu, and C.-Z. Chang, "Crossover from Ising- to Rashba-type superconductivity in epitaxial Bi2Se3/monolayer NbSe2 heterostructures," Nat. Mater. **21**(12), 1366–1372 (2022).
[48] A. Kudriashov, X. Zhou, R.A. Hovhannisyan, A.S. Frolov, L. Elesin, Y.B. Wang, E.V. Zharkova, T. Taniguchi, K. Watanabe, Z. Liu, K.S. Novoselov, L.V. Yashina, X. Zhou, and D.A. Bandurin, "Non-Majorana origin of anomalous current-phase relation and Josephson diode effect in Bi2Se3/NbSe2 Josephson junctions," Sci. Adv. **11**(24), eadw6925 (2025).
[49] E. Wang, H. Ding, A.V. Fedorov, W. Yao, Z. Li, Y.-F. Lv, K. Zhao, L.-G. Zhang, Z. Xu, J. Schneeloch, R. Zhong, S.-H. Ji, L. Wang, K. He, X. Ma, G. Gu, H. Yao, Q.-K. Xue, X. Chen, and S. Zhou, "Fully gapped topological surface states in Bi2Se3 films induced by a d-wave high-temperature superconductor," Nat. Phys. **9**(10), 621–625 (2013).
[50] V. Lahtinen, and J.K. Pachos, "A Short Introduction to Topological Quantum Computation," SciPost Phys. **3**(3), 021 (2017).
[51] D.H. Ahn, J. Garlick, M. Grondona, D. Lipari, B. Springmeyer, and M. Schulz, "Flux: A Next-Generation Resource Management Framework for Large HPC Centers," in *2014 43rd Int. Conf. Parallel Process. Workshop*, (IEEE, Minneapolis, MN, USA, 2014), pp. 9–17.
[52] A. Zur, and T.C. McGill, "Lattice match: An application to heteroepitaxy," J. Appl. Phys. **55**(2), 378–386 (1984).
[53] Y. Cao, V. Fatemi, S. Fang, K. Watanabe, T. Taniguchi, E. Kaxiras, and P. Jarillo-Herrero, "Unconventional superconductivity in magic-angle graphene superlattices," Nature **556**(7699), 43–50 (2018).
[54] C. Zhang, C.-P. Chuu, X. Ren, M.-Y. Li, L.-J. Li, C. Jin, M.-Y. Chou, and C.-K. Shih, "Interlayer couplings, Moiré patterns, and 2D electronic superlattices in MoS2/WSe2 hetero-bilayers," Sci. Adv. **3**(1), e1601459 (2017).
[55] K. Momma, and F. Izumi, "VESTA: a three-dimensional visualization system for electronic and structural analysis," J. Appl. Crystallogr. **41**(3), 653–658 (2008).
[56] S. Grimme, J. Antony, S. Ehrlich, and H. Krieg, "A consistent and accurate ab initio parametrization of density functional dispersion correction (DFT-D) for the 94 elements H-Pu," J. Chem. Phys. **132**, 154104 (2010).

[57] E. Gerber, S.B. Torrisi, S. Shabani, E. Seewald, J. Pack, J.E. Hoffman, C.R. Dean, A.N. Pasupathy, and E.-A. Kim, "High-throughput ab initio design of atomic interfaces using InterMatch," Nat. Commun. **14**(1), 7921 (2023).
[58] J.P. Heremans, R.J. Cava, and N. Samarth, "Tetradymites as thermoelectrics and topological insulators," Nat. Rev. Mater. **2**(10), 17049 (2017).
[59] Y. Ren, Y. Li, and X. Liu, "Investigation on band alignment of Bi2Se3–PbSe heterojunction," Appl. Phys. Lett. **118**(16), 162101 (2021).
[60] S. Yi, C. Chen, M. Yu, J. Hao, and Y. Wang, "2D/2D Bi2Se3/SnSe2 heterostructure with rapid NO2 gas detection," Front. Chem. **12**, (2024).
[61] I.V. Antonova, N.A. Nebogatikova, N.P. Stepina, V.A. Volodin, V.V. Kirienko, M.G. Rybin, E.D. Obrazstova, V.A. Golyashov, K.A. Kokh, and O.E. Tereshchenko, "Growth of Bi2Se3/graphene heterostructures with the room temperature high carrier mobility," J. Mater. Sci. **56**(15), 9330–9343 (2021).
[62] X. Jiang, T. Yilmaz, E. Vescovo, and D. Lu, "Manipulating topological properties in Bi2Se3/BiSe transition metal dichalcogenide heterostructures with interface charge transfer," Phys. Rev. B **109**(11), 115112 (2024).
[63] K.H. Yeoh, K.H. Chew, T.L. Yoon, Rusi, Y.H.R. Chang, and D.S. Ong, "Metal to semiconductor transition of two-dimensional NbSe2 through hydrogen adsorption: A first-principles study," J. Appl. Phys. **128**(10), 105301 (2020).

# IX. Appendix

## Appendix A: Implementation Details

*Configuring User Settings with setup_config*

The `config_pyrmg` executable configures system settings for supercomputing environments, such as resource allocation, default partition auto-generation, and the location of the RMG binary. By default, the config.yml configuration YAML is stored in the user's `~/.pyRMG/` folder, enabling customization for each individual user even when RMG or pyRMG are installed in a shared location or environment (e.g., via Conda, Python environment, or Spack).

*Generating RMG Input Files with generate_pyrmg*

With pyRMG, RMG input files are generated by the `generate_pyrmg` executable, which requires two passed arguments: (i) a structure file name (currently supported formats include a VASP POSCAR or .cif files, an RMG log file from a previous run, or an existing RMG input file) and (ii) an input YAML file name containing RMG options listed on the RMG GitHub Wiki at RMG Input-File Options, as well as some custom tags: these include the "cutoff" and "kdelt" tags that can be passed in the input YAML autogenerate the "wavefunction_grid" and "kpoint_mesh" options following the same convention as RMG's GUI backend. Parameters in the YAML configuration file (specified by the `--rmg_yaml` flag) are applied uniformly to all eligible structure files in the directory indicated by the `--parent_directory` flag.

Additional tags that have no corollary to the RMG GUI backend include the "unoccupied_fraction", which sets the number of unoccupied orbitals for the calculation based on the number of valence electrons used by the pseudopotentials, "per_atom_energy", which sets the “energy_convergence_criterion” flag based on the total number of electrons in the calculation, and the “per_atom_rms”, which sets the “rms_convergence_criterion” based on the total number of electrons in the calculations. These user-defined tags supersede certain control options defined on the Wiki that are accepted as RMG inputs (e.g., “unoccupied_states_per_kpoint”, “energy_convergence_criterion”, and “rms_convergence_criterion”).

*Job Submission Template*

`generate_pyrmg` also accepts a job submission template (via the `--rmg_submission` flag) that defines the modules to be loaded and resources to be allocated for each run. The job submission template should be customized according to the system configuration. An example submission script for Frontier is provided in the pyRMG submission_templates directory, where placeholders (e.g., `{ALLOCATION}`, `{TIME}`, `{NODES}`) are substituted with user-specified values and copied to directories with structure files during the execution of `generate_pyrmg`.

*Single-Job Submissions with submit_pyrmg*

Individual input files generated by `generate_pyrmg` can be submitted using the `submit_pyrmg` executable with the `--parent_directory` and `--submit` flags; without the `--submit` flag, `submit_pyrmg` only reports the status of jobs. `submit_pyrmg` navigates the specified `--parent_directory`, checks for the presence of an appropriate RMG input and submission script, and submits the job using `sbatch` if the job has not yet converged. For “Quench Electrons” jobs, only SCF convergence is checked; for “Relax Structure” jobs, both SCF and force convergence must be achieved. `generate_pyrmg` does not regenerate inputs for, and `submit_pyrmg` does not resubmit, jobs that have already converged.

*Processor Grid Assignment Algorithm:*

Because RMG is GPU-accelerated, pyRMG's auto-grid approach currently only targets GPU-aware RMG builds (e.g., Frontier, Perlmutter, Aurora). The routine constructs candidate 3-D processor grids $\{p_x, p_y, p_z\}$ and selects the grid that balances three conditions:

a. Compatibility with RMG's finite difference constraints (controlled by the --grid_divisibility_exponent, $n$)
b. The per-calculation resources the user is targeting (controlled by --electrons_per_gpu), and
c. Domain homogeneity, or an even distribution of the real-space wavefunction grid (based on unit cell) workload across processor subdomains.

We detail the approach below and define the scoring function used to pick the optimal candidate grid.

1. Build candidate grids

pyRMG first forms a candidate grid, $\{p_x,\ p_y\ , p_z\}$, from processors (GPUs) distributed along the x, y and z directions. The total processors used, $\mathrm{G}_{cand}$, is:

$$\mathrm{G}_{cand} = p_x \times\ p_y\ \times\ p_z \qquad (A1)$$

2. Grid divisibility (hard filter)

pyRMG requires that the total processors meets the finite-difference stencil requirement based on the user's `--grid_divisibility_exponent`, $n$. $G_{cand}$ not divisible by $2^n$ are discarded.

3. Score remaining candidates

Each candidate grid is scored based on the user's requested resources and the domain homogeneity of the candidate grid. We first compute an *ideal processor count* from the total valence electrons in the cell:

$$G_{ideal} = \frac{Total\ valence\ electrons}{Electrons\ per\ GPU\ (user\ flag)} \quad (A2)$$

Two components contribute to the score:

**GPU penalty**: this captures how far the candidate grid's GPU count is from the ideal processor count:

$$GPU\ Penalty = \alpha \cdot \frac{|G_{cand} - G_{ideal}|}{G_{ideal}} \quad (A3)$$

To bias against undersized grids (which risk out-of-memory failures), pyRMG uses a larger weight for candidates that allocate fewer GPUs than the ideal. By default:

$$\alpha = \begin{cases} 0.2, & G_{cand} < G_{ideal} \\ 0.1 & otherwise. \end{cases} \quad (A4)$$

Users can tune α based on their workflow needs.

**Heterogeneity penalty**: this measures how uneven the per-processor, real-space workload would be. Let the wavefunction grids along each cartesian direction be $\{w_x,\ w_y\ , w_z\}$, and the candidate grid be $\{p_x,\ p_y\ , p_z\}$. The wavefunction grid per-processor (per-PE) is:

$$per - PE = \left(\frac{w_x}{p_x}, \frac{w_y}{p_y}, \frac{w_z}{p_z}\right) \quad (A5)$$

And its standard deviation is $\sigma = std(per - PE)$. To avoid $\sigma = 0$ for perfectly uniform $per - PE$, we impose a small floor (tolerance = 0.1) and define:

$$Heterogeneity\ Penalty = \max(\sigma, tolerance) \quad (A6)$$

The final score is the produce of the two penalties, with lower scores indicating better candidates:

$$score = GPU\ Penalty\ \times\ Heterogeneity\ Penalty \quad (A7)$$

4. Final GPU count and k-point parallelism

The optimal grid defines the total GPUs for the configuration,

$$G_{conf} = G_{cand}^{(best)}, \min(score) \quad (A8)$$

If k-point parallelization is used, pyRMG multiplies the configuration GPUs by the k-point distribution factor

$$\mathrm{G}_{total} = \mathrm{G}_{conf} \times\ \mathrm{G}_{kpoints} \quad (A9)$$

Where $\mathrm{G}_{kpoints}$ is either the user's "kpoint_distribution" parameter or an automatic choice (pyRMG follows RMG's default behavior when "kpoint_distribution = -1".)

*Processor Grid Selection: Nb (110)*

To illustrate the choice of processor grid, we consider the Nb (110) slab with six Nb atoms (78 valence electrons) and 20 Å of vacuum space (lattice parameters a = 3.259 Å, b = 4.609 Å, and c = 27.654 Å). The user-specified "cutoff" parameter, equal to 300 Rydberg, generates a wavefunction grid of [34, 48, 288]. This cutoff is higher than necessary to converge the current Nb (110) slab, but it makes the processor grids more variable for this example. Setting `--electrons_per_gpu=5` and `--grid_divisibility_exponent=1`, pyRMG first constructs a default processor grid by normalizing the wavefunction grid by its minimum component ($\frac{34}{34}, \frac{48}{34}, \frac{288}{34}$) and flooring. This results in a default processor grid of [1, 1, 8], which scores 2098.38.

Next, given the 78 electrons in the cell, and the target --electrons_per_gpu=5, pyRMG reduces the smallest component of the processor grid in integer steps from

$$\left\lceil \left(\frac{78}{5}\right)^{1/3} \right\rceil = 3 \; down \; to \; 1 \qquad (A10)$$

recomputing $\{p_x, p_y, p_z\}$ and updating the best score at each step. For $p_x = 3$, the default grid remains the most optimal; for $p_x = 2$, the best grid becomes [2, 3, 17] with a score of 368.87; and for $p_x = 1$, the best grid becomes [1, 2, 9] with a score of 66.97. The final wavefunction grid uses $1 \times 2 \times 9 = 18$ GPUs, corresponding to $78/18 \approx 4.33$ electrons per GPU.

*Bulk Relaxations of $Bi_2Se_3$ and $NbSe_2$*

Before constructing the $Bi_2Se_3$/$NbSe_2$ heterostructures, we relaxed the cells and ionic positions of the bulk, van der Waals-layered $Bi_2Se_3$ and $NbSe_2$ structures. Figure 10 shows the submission script containing our workflow:

1. Activate a Conda environment where pyRMG is installed.
2. Run `config_pyrmg` (if needed) to generate ~/.pyRMG/config.yml with HPC-specific settings (e.g., allocation, partition, or executable location).
3. Run `generate_pyrmg` to scan the directories containing $Bi_2Se_3$ and $NbSe_2$ structures for unconverged jobs and generate new rmg_input files from the YAML configuration and job-template files, using the settings in ~/.pyRMG/config.yml.
4. Run `submit_pyrmg` from the same directory with the --submit flag specified to submit a series of individual RMG jobs.

```
#-------------- SETUP AND SUBMISSION -----------
#
#Step 1: Generate config.yml with your settings
config_pyrmg --allocation {$ALLOC} --partition ...
#
#Step 2: Generate rmg_input submission files
generate_pyrmg -pd {$RUN_PATH} -ry {$YAML_FILE} ...
#
#Step 3: Submit jobs individually
submit_pyrmg -pd {$RUN_PATH} --submit
```

**Figure 10.** Shell script containing the pyRMG executables needed to auto-generate inputs for and submit individual RMG calculations with convergence checking built-in.

This workflow also addresses a common failure of RMG "Relax Structure" jobs when "cell_relax = True": as the lattice relaxes, it might no longer be commensurate with the original wavefunction grid or k-point mesh, causing the calculation to abort. While pyRMG does not prevent this failure, when regenerating the grid and mesh for unconverged jobs, `generate_pyrmg` will first check the .log files for a valid structure, then any existing rmg_input files, and finally the original structure (e.g., POSCAR). This means that inputs are regenerated from the last valid structure before job failure (because of the grid, due to wall-time, etc.) without the user having to extract the correct image for job continuation manually.

For this example, bulk $Bi_2Se_3$ converged in 27 ionic-cell optimization steps without any restarts, whereas bulk $NbSe_2$ required four resubmissions due to grid mismatches. Because pyRMG's submission scripts automate resource allocation, input-file generation, convergence checks, structure and grid updates, and automatic resubmission, we only needed to rerun the shell script four times to reach full convergence. The single-job framework requires no additional installation and provides a robust backbone for routine calculations. Single-job submission was used for ion-optimized and SOC calculations in our study of 2L-$Bi_2Se_3$/2L-$NbSe_2$ heterostructures.

*Submission of RMG jobs via pyRMG and MatEnsemble*

Perhaps more importantly, pyRMG's core features can be embedded into custom workflows and specialized environments to enable true high-throughput investigations of challenging systems. To demonstrate this, we developed `matsemble_pyrmg`, an executable in pyRMG that integrates with the MatEnsemble framework. This framework utilizes the executor interface of the Flux job scheduler and implements an adaptive HPC resource management to maximize the throughput on extreme-scale computing platforms (e.g. Frontier). Figure 11 shows our master submission script, which uses a Conda environment containing MatEnsemble and pyRMG along with a Spack build containing Flux. Rather than submitting calculations serially with `submit_pyrmg`, `matsemble_pyrmg` requests a pool of resources and dynamically launches new asynchronous RMG calculations as other jobs finish or fail, continuing until the full workflow has converged or the wall-time clock limit is reached.

Module loads, sbatch flags, and environment variables are included in the master submission script and broadcast to each task, while per-job requirements (i.e., nodes, GPUs) based on auto-generated processor grids are still read from single-job scripts in each subdirectory, utilizing the file generation logic from Figure 10. Integrating pyRMG into a workflow scheduler is particularly important for large jobs that request a large number of nodes but complete at various times before a system-specific wall time, i.e., single point calculations of heterostructures with variable supercells. Multi-job submission with MatEnsemble was used for SCF calculations in our study of 2L-$Bi_2Se_3$/2L-$NbSe_2$ heterostructures.

```
#!/bin/bash
#
#---------------- SBATCH ARGUMENTS ---------------
# e.g., nodes requested for MatEnsemble + Flux
#SBATCH -N ...
#
#---------------- LOAD RMG MODULES ---------------
#
# HPC-specific RMG modules
module load example/1.2.3
#
# Environment variables
export VAR1=
#
# --------------- ACTIVATE pyRMG CONDA IN SHELL --
#
conda activate {$CONDA_PATH}
#
# --------------- LOAD SPACK ---------------------
#
. {$SPACK_PATH}
spack env activate {$SPACK_NAME}
spack load flux-sched
#
# --------------- ACTIVATE pyRMG CONDA IN SPACK --
#
conda activate {$CONDA_PATH}
#
#--------------- SETUP AND SUBMISSION ------------
#
#Step 1: Generate config.yml with your settings
config_pyrmg --allocation {$ALLOC} --partition ...
#
#Step 2: Generate rmg_input submission files
generate_pyrmg -pd {$RUN_PATH} -ry {$YAML_FILE} ...
#
#Step 3: Submit jobs in parallel with MatEnsemble
srun --external-launcher --mpi=pmi2 \
     --gpu-bind=closest flux start \
     matsemble_pyrmg -pd {$RUN_PATH} ...
```

**Figure 11.** Master submission script for pyRMG heterostructure workflow, with examples that can be found in the GitHub. Key differences between the single-job shell script of Figure 10 include the specification of RMG job setups in the submission script (broadcast to individual tasks in the full workflow), Conda and Spack loading, and execution of `matsemble_pyrmg` instead of `submit_pyrmg` to run tasks dynamically.

## Appendix B: Tests and Preliminary Results

*Exfoliation Energy Curve*

To create the exfoliation energy curves, we computed single-point calculations for two model 2L-$Bi_2Se_3$/2L-$NbSe_2$ systems (54 and 234 atom cells) at 8 interface distances (2.406, 2.539, 2.606, 2.673, 2.74, 2.94, 3.208, and 4.01 Å.) Based on the exfoliation energy curves, we fixed the interface distance to 3.2 Å for our investigation.

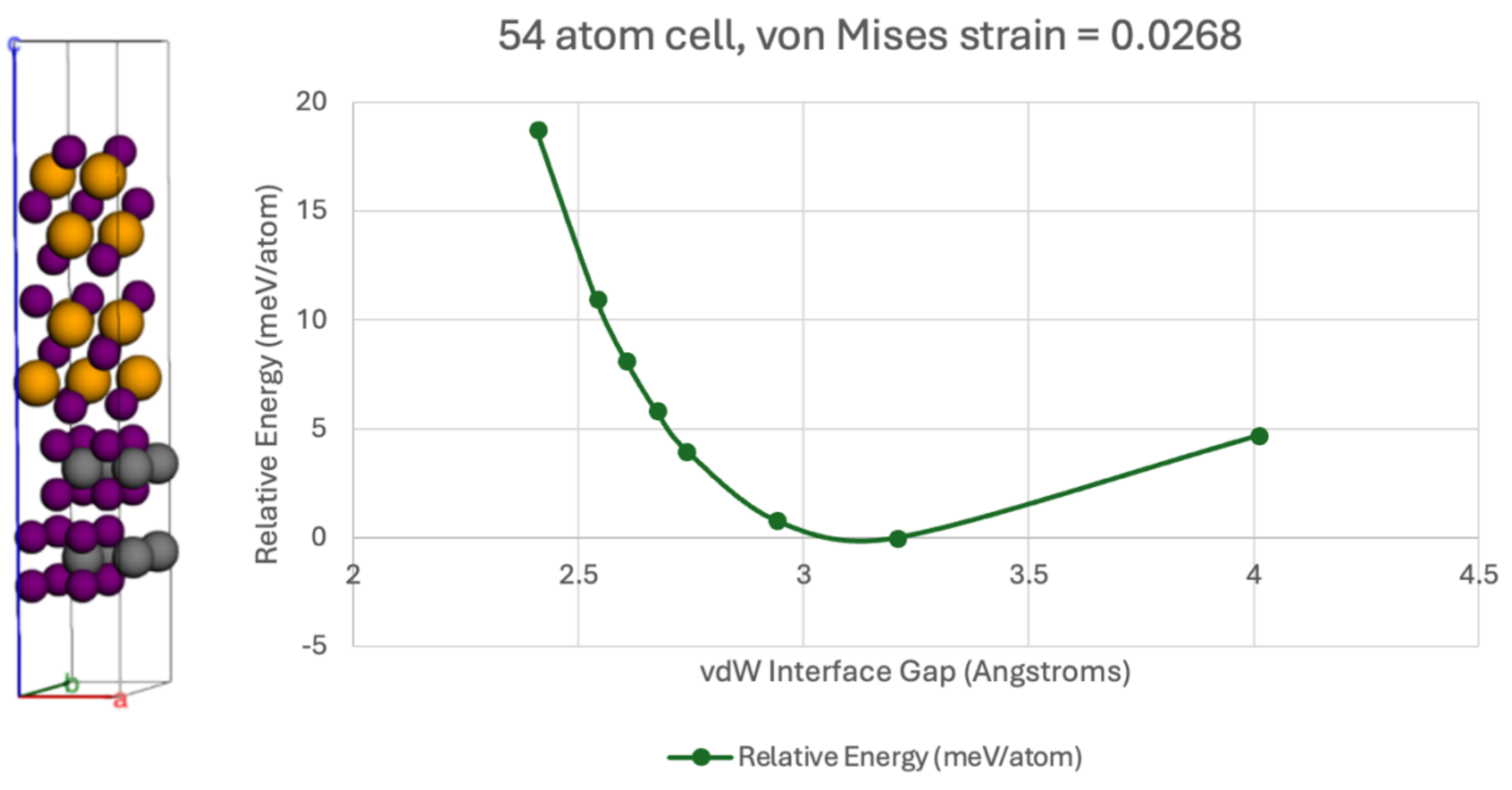

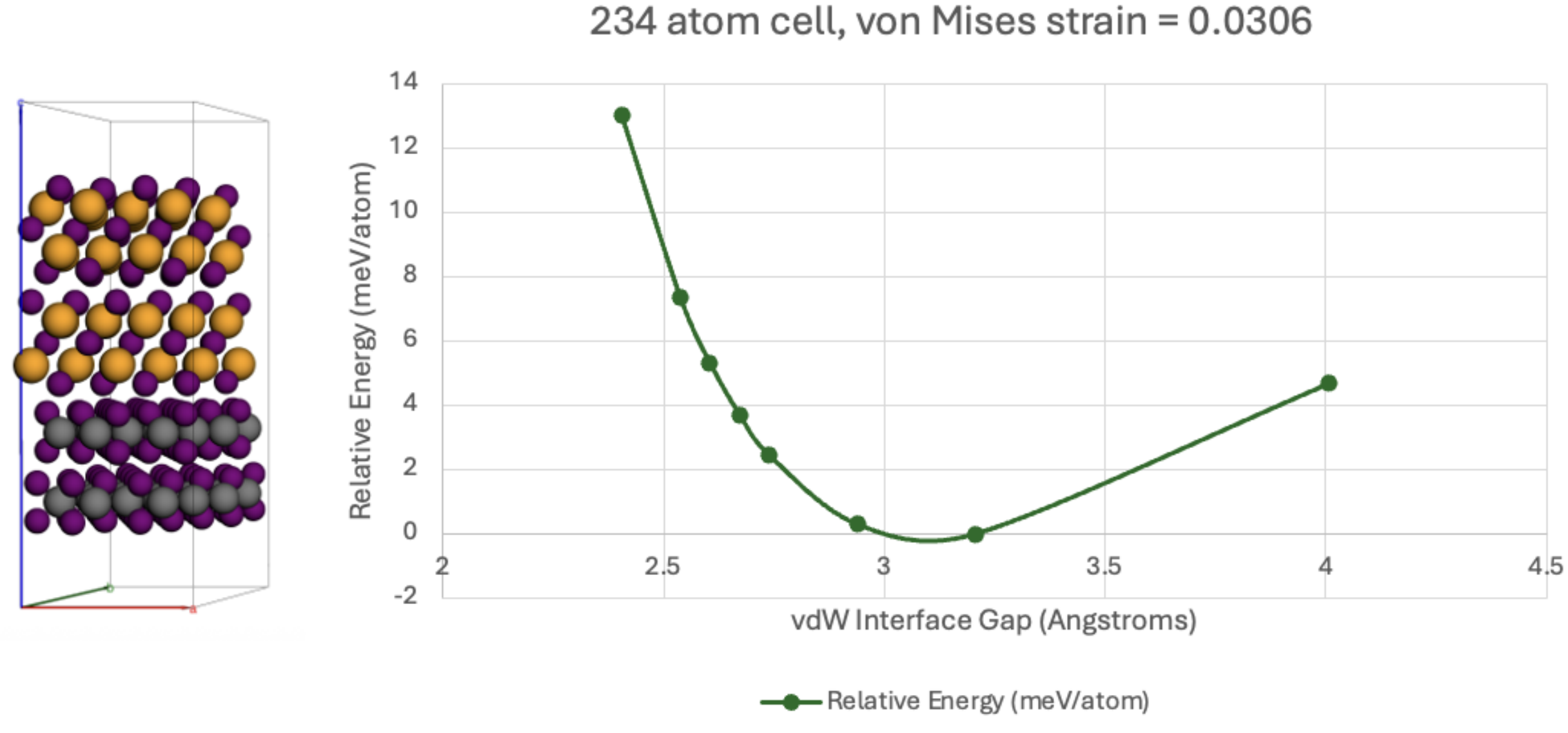

**Figure 12.** Exfolation energy curves for (top) 54-atom and (bottom) 234-atom 2L-$Bi_2Se_3$/2L-$NbSe_2$ heterostructures, where the dots represent structures where DFT energies were computed and the line is a fitted interpolation between points. The minimum energy structures for both systems lies between sampled points 2.94 Å and 3.208 Å: we would expect slight differences in exact minimum spacing in each individual system based on the strain and twist angle, defining coordination at the interface, but for SCF calculations in this investigation we fixed the interface distance to 3.2 Å.

*Davidson vs. Multigrid Solver Benchmarking*

We benchmarked SCF step timings for the Davidson and multigrid solvers using the 168-atom and 656-atom heterostructures, which had been ionically optimized as described in the "Ionic Relaxations" section of the main text. For the 168-atom system (4 k-points, 944 wavefunctions), the multigrid solver does not become more efficient than Davidson until GPU counts exceed roughly 1,600; at GPU counts below ~500, which were more representative of our high-throughput workflow, the Davidson solver remains substantially faster. In contrast, the multigrid solver outperforms Davidson at all GPU counts tested for the 656-atom system (1 k-point, 3,680 wavefunctions). For both systems, multigrid SCF step times continue to decrease as GPU count increases, while Davidson timings plateau due to communication overhead.

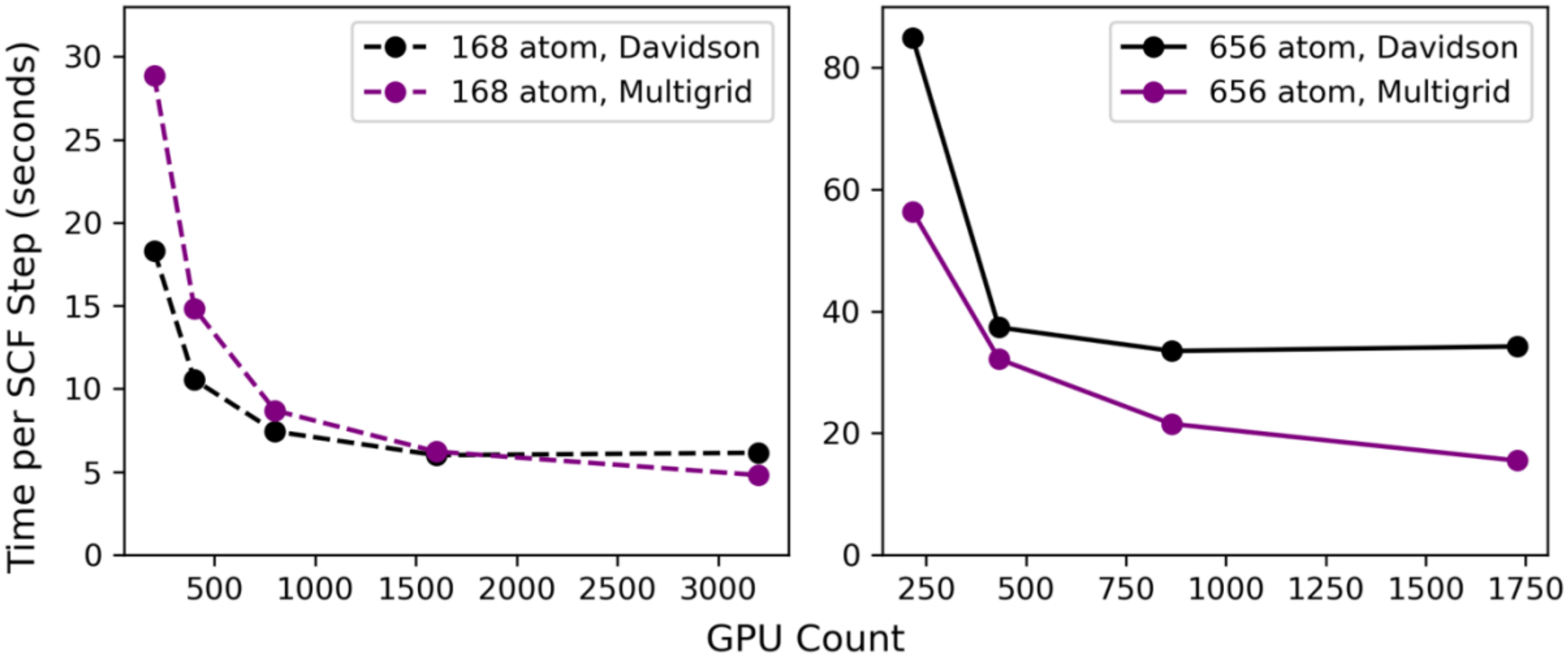

**Figure 13.** Time per SCF step as a function of GPU count for the Davidson and multigrid solvers for the (left) 168-atom and (right) 656-atom heterostructures included in our high-throughput workflow.

*Preliminary Results: Folding the Twist Angles*

We first performed a high-throughput DFT single point investigation on 91 structures generated with HeteroBuilder: we loosened the von Mises strain requirement for these structures, so that the maximum was solved to be 0.069, and limited ourselves to heterostructures with 422 atoms or fewer. The vacuum spacing for heterostructures in this dataset was 15.000 Å, and the interface spacing was 2.673 Å, which were created prior to generating the exfoliation energy curve. Compared to our dataset reported in the main text, this preliminary dataset has a broader range of von Mises strains and a more narrow range of total atom counts.

Figure 14 (left) shows the formation energies of 56 structures, which have been grouped based on their modulo 60° twist angles, $\theta_{mt}$ (structures with the lowest atom counts taken as representatives). In addition to their modulo 60° twist angles, we also mirror these values around 30° ($\theta_{rt}$) which (most notably) result in similar formation energies for the low-energy twist structures (purple). Accounting for this mirroring with $\theta_{rt}$ in Figure 14 (right) results in formation energies that are strongly, positively correlated with the von Mises strain, with Pearson coefficient r = 0.740.

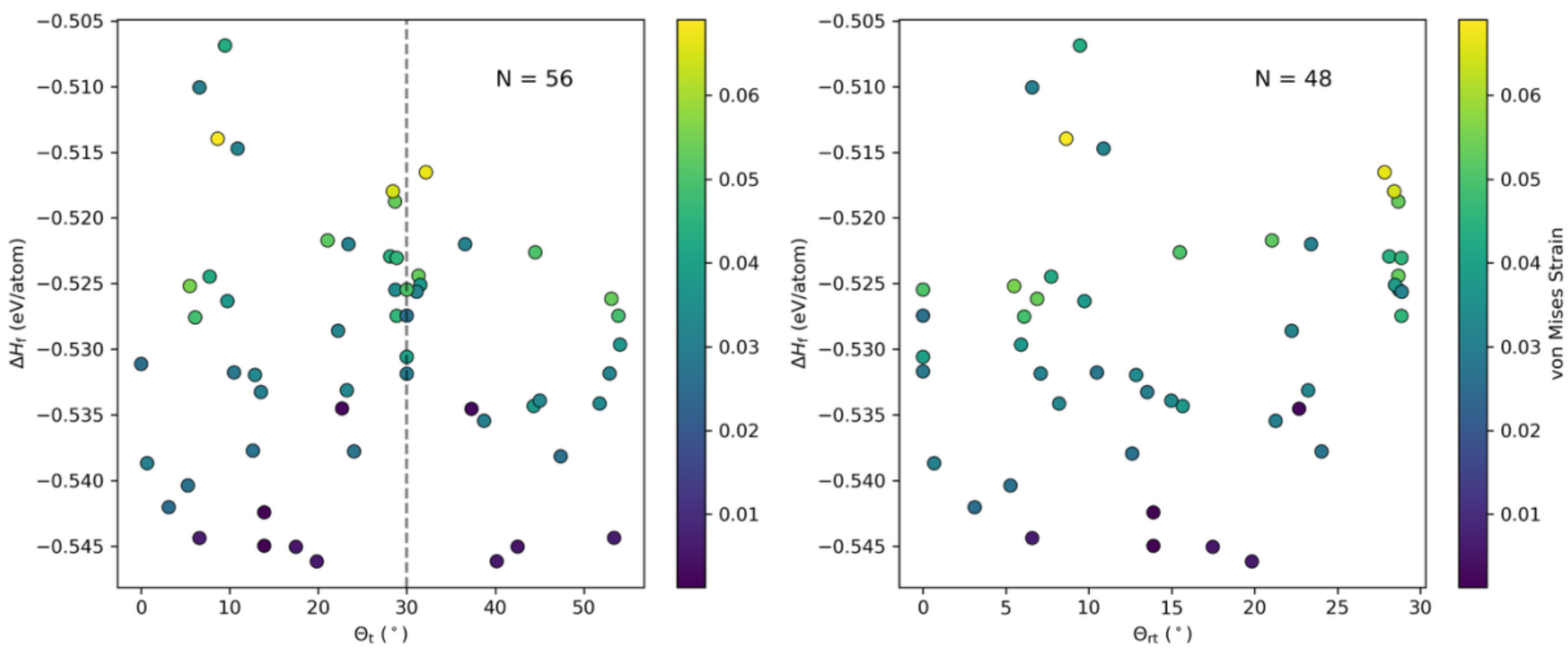

**Figure 14.** Formation energy relationships for the (a) 56 structures chosen as representatives from the full dataset of 91 heterostructures, based on their modulo 60°**,** $\theta_{mt}$, values, and (b) the 48 heterostructures reduced by mirroring the systems around 30°.

*Preliminary Results: Wavefunction Grid Anisotropy*

Figure 15 (left) plots the time required to converge SCF steps for all 91 heterostructures in the preliminary dataset (x-axis) to the electrons per GPU solved for by the optimal solution to the processor grid (y-axis) and the processor grid anisotropy, $\mathcal{A}$ (color bar). `electrons_per_gpu` = 4 was set when generating inputs to run the high-throughput workflow for this dataset, and all but 3 calculations were converged with

`kpoint_distribution` = 1. Most of the calculations (47/91) have processor grids that solve for 3 < electrons_per_gpu < 5 and SCF time < 30 minutes, showing that greater than 50% of jobs use close to the resources specified by the user and finish rapidly.

Runtimes in Figure 15 (left) are more strongly correlated with $\mathcal{A}$ (r = 0.834) than with the number of atoms (right, r = 0.128), though the maximum total number of atoms in any calculation in this is 422 and this correlation does not hold for the data reported in the main text, where the maximum number of atoms is 1,046. 68 runs finish within 30 minutes having $\mathcal{A}$ < 4.06, while max($\mathcal{A}$) = 5.73. We attribute the long tail of slower runtimes for calculations that take more than 30 minutes to finish to large processor grid anisotropies, which can limit the efficiency of workload management across the distributed processors.

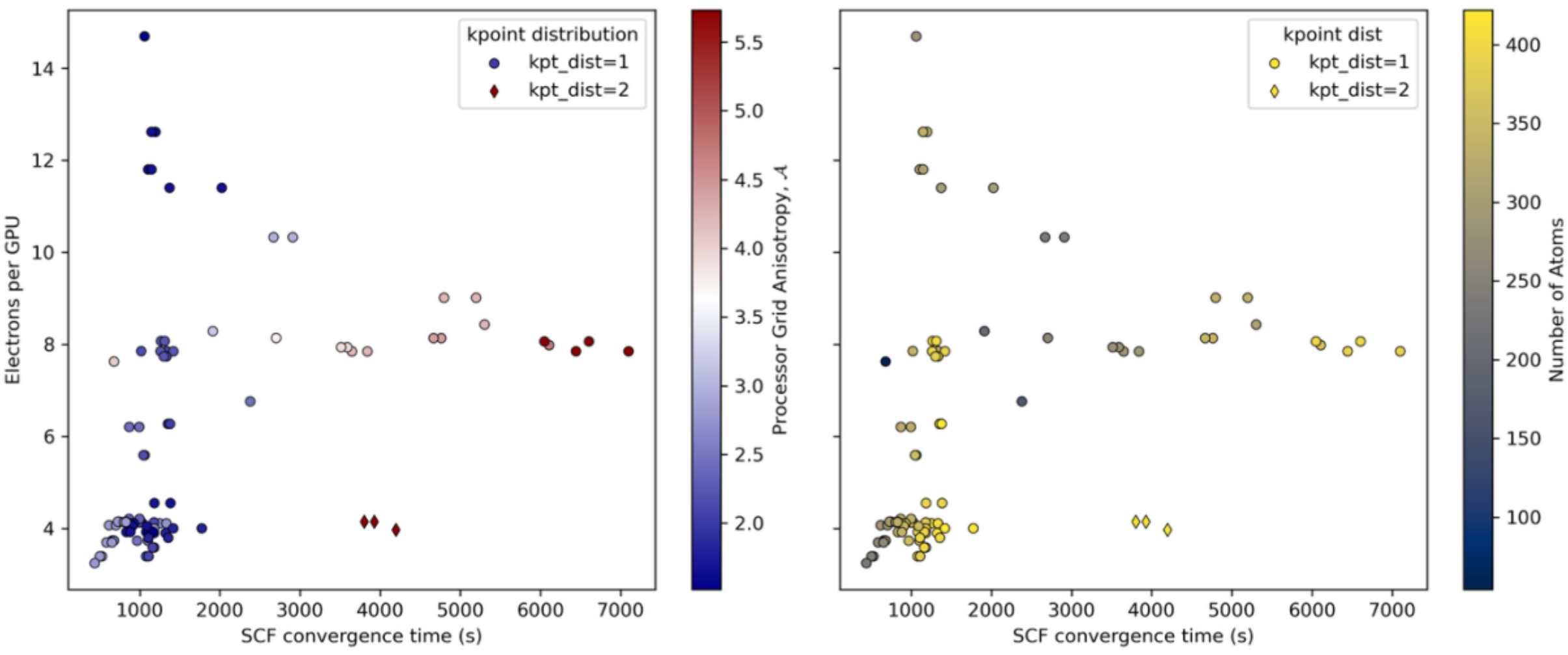

**Figure 15.** Single-point calculation convergence time vs electrons per GPU with color bars indicating (left) wavefunction grid anisotropy, $\mathcal{A}$, and (right) the number of atoms in the cell for the 91 2L-$Bi_2Se_3$/2L-$NbSe_2$ heterostructures generated with Zur's algorithm in our preliminary dataset. Calculation times for the full dataset are more strongly correlated with $\mathcal{A}$ (pearson coefficient, ρ = 0.837) than atom number (ρ = 0.128), demonstrating that RMG scales nearly linearly with respect to electrons per GPU for these 2D systems with similar grid anisotropies.